# AI-Assisted Writing is Growing Fastest among Less Established Scientists in Non-English-Speaking Countries

**Authors:** Jialin Liu[1], Yongyuan He[2], Zhihan Zheng[3], Yi Bu[2, *], Chaoqun Ni[1, *]

[1] Information School, University of Wisconsin-Madison, Madison, WI 53706, U.S.A.

[2] Department of Information Management, Peking University, Beijing 100871, China

[3] Faculty of Education, The University of Hong Kong, Hong Kong 999077, China

* Corresponding Authors. Email: buyi@pku.edu.cn and chaoqun.ni@wisc.edu.

## Abstract

The recent emergence of AI-assisted writing raises an important question: how is this new technology being adopted across the scientific community, and how does adoption vary across linguistic and professional contexts? We analyze over two million full-text biomedical publications from PubMed Central from 2021 to 2024 using a distribution-based framework to estimate AI-generated content. We found that, in biomedical publications, AI-generated content increased substantially after ChatGPT, with larger increases in publications from countries with lower English proficiency. Increases were also greater among scientists with fewer publications and citations, those at earlier career stages, and those at lower-ranked institutions. Prior AI research experience was associated with greater increases in AI-assisted writing, which were also modestly associated with greater increases in publication productivity. These findings show that AI-assisted writing is growing fastest among biomedical scientists who may have historically faced barriers, a pattern with potentially positive implications for equity in science.

## Introduction

Generative artificial intelligence (GenAI), particularly large language models (LLMs), is rapidly becoming part of scientific writing. Researchers report using these tools across a growing range of scholarly tasks [1–5], and large-scale analyses of published articles show a marked increase in linguistic signatures associated with AI-assisted writing following the release of ChatGPT [6–12]. This emerging literature has established that AI-assisted writing is spreading through science. Far less is known, however, about how that diffusion is distributed across the scientific workforce. New technologies rarely diffuse uniformly: their adoption depends not only on availability, but also on the benefits users expect from them, their compatibility with existing practices, the skills required to use them, and the environments in which adoption takes place [13–15]. The rapid entry of LLMs into scientific writing therefore raises an important question that aggregate estimates of LLM use cannot answer: who is adopting AI-assisted writing more, and how does adoption vary across national linguistic, professional, and institutional contexts? Because the AI-assisted writing examined here is primarily enabled by LLM-based tools such as ChatGPT and Gemini, we use "LLM" for the underlying writing technology and "GenAI" for the broader class of generative technologies.

The question above is particularly consequential because researchers differ substantially in the linguistic, professional, and institutional constraints they face in scientific writing. English has become the dominant language of international scientific communication [16,17], but producing publication-ready English places substantial additional demands on researchers working in contexts where English proficiency is lower [18–23]. LLM-based writing tools may lower some of these costs by assisting with drafting, editing and linguistic refinement, and may therefore be especially valuable where language-related barriers are greater. Furthermore, researchers differ in accumulated scholarly experience and visibility, career stage, access to institutional resources and familiarity with AI technologies. These differences may shape the perceived benefits of AI-assisted writing, its compatibility with existing work practices, and ultimately its adoption patterns. Uneven diffusion could therefore have important distributional consequences: if adoption is concentrated among already advantaged researchers, it could reinforce existing disparities; if it occurs more rapidly among researchers facing greater linguistic, professional, or institutional constraints, it could instead have compensatory potential. Yet greater access to writing assistance does not necessarily translate into greater equality. Concerns remain about authorship and disclosure, the homogenization of scientific expression, and the broader consequences of GenAI for scholarly communication[12,24–29]. Understanding who adopts AI-assisted writing, and under what conditions, is therefore a necessary step toward assessing what its diffusion may ultimately mean for inequality in science.

Existing studies have begun to document heterogeneity in AI-assisted writing across countries, publication venues, and disciplines[9,30–33]. Yet these findings remain largely disconnected from the characteristics of the researchers and institutions producing the publications. Many observational studies examine publications or aggregate groups, while relatively little is known about whether differences in adoption extend systematically across researchers' scholarly trajectories, career stages, institutional positions, and prior technological experience. The key empirical gap is therefore not simply whether AI-assisted writing varies across settings, but whether these differences constitute a broader pattern of uneven diffusion across researchers occupying different linguistic, professional, and institutional positions in science.

To address this gap, we analyze more than two million full-text biomedical publications from the PubMed Central (PMC) Open Access Subset between 2021 and 2024. Using a distribution-based framework developed by Liang *et al.* [32], we estimate changes in AI-assisted writing and examine how they vary across national linguistic environments, authors' cumulative productivity and citation impact, career seniority, and institutional prestige (Methods). We further test whether prior AI research experience is associated with greater increases in AI-assisted writing and whether changes in AI-assisted writing are associated with changes in publication productivity. Country assignment is based on authors' institutional affiliations and is used to characterize the national linguistic environment in which researchers work rather than their individual native language or English proficiency. By linking large-scale full-text evidence with longitudinal author and institutional characteristics, this study moves beyond documenting the presence of AI in scientific writing to characterize the social patterning of its diffusion across the biomedical research community.

## Results

### Greater Increase in AI-generated Content in Papers from Non-English-Speaking Countries

Our analysis reveals substantial cross-country variation in the post-ChatGPT increase in AI-assisted writing, with the magnitude of this increase systematically associated with countries' linguistic characteristics. Fig. 1a shows the fraction of AI-generated sentences in biomedical publications from English- and non-English-speaking countries from 2021 to 2024. Prior to the release of ChatGPT (2021–2022), the estimated use of AI-assisted writing remained relatively stable across both groups, with English-speaking countries showing a slightly higher fraction. However, a significant shift occurred after the release of ChatGPT (2023–2024). While both groups demonstrated a significant upward trajectory, publications from non-English-speaking countries witnessed AI-generated content escalate dramatically from 0.038 to 0.193, a growth rate considerably faster than that observed in English-speaking countries, where the fraction increased from 0.057 to 0.171. Results from the difference-in-differences (DiD) analysis (Methods) further support this observation: After controlling for other covariates, our study reveals a significant differential change (Fig. S1 and Table S2): the fraction of AI-generated sentences in publications from non-English-speaking countries increased by an average of 0.02 units (i.e., the fraction of AI-generated sentences) ($p < .001$) compared to publications from English-speaking countries after the release of ChatGPT. These results provide statistical evidence that the adoption of AI-assisted writing differs systematically across national linguistic contexts.

Building on our observation of a broader split between English-speaking and non-English-speaking countries, we next conducted a finer-grained, country-level analysis to explore the specific relationships between AI adoption, geography, and English proficiency. For each country, we calculated the change in the average fraction of AI-generated content between 2021–2022 and 2023–2024 (Fig. 1c). Our findings indicate that publications originating from countries in Asia, North Africa, and South America demonstrate a higher increase in AI-generated content compared to those from Northern and Western Europe and North America. This geographic pattern became clearer when combined with linguistic metrics, revealing a strong negative correlation (Spearman's $\rho$ = -.652, $p < .001$) between a country's increase in AI-assisted writing and its English Proficiency Index (EPI) (Fig. 1b), which measures the English-language proficiency of a country's adult population [34] (Methods; Supplementary Note 1). In short, while AI-assisted writing increased in all countries, the *magnitude* of this increase was inversely related to a country's English proficiency. The most significant increases, therefore, are concentrated in countries with both lower English proficiency and greater linguistic distance from English. For example, while the average fraction of AI-generated content in publications from the Netherlands (EPI Rank = 1st, Germanic family) increased by 0.035, that in publications from China (EPI Rank = 91st, Non-Indo-European family) increased by 0.102. These patterns indicate that post-ChatGPT increase in AI-assisted writing was systematically greater in national contexts with lower English proficiency and greater linguistic distance from English.

## Author-Level Heterogeneity in AI Adoption in Academic Writing

Having observed substantial differences in AI adoption across national linguistic environments, we next examined whether adoption patterns also vary across author profiles. To do so, we stratified publications according to authors' productivity (top 5% vs. non-top 5% by their cumulative number of publications), citation impact (top 5% vs. non-top 5% by their cumulative number of citations), career stage (junior authors being those with less than 15 years of publication history vs. senior authors), and institutional prestige (top 100 vs. non-top 100 institutions by 2025 QS World University Rankings in Life Sciences and Medicine). We then estimated the changes from pre- (2021-2022) to post-ChatGPT (2023-2024) using difference-in-difference-in-differences (DDD) models (Methods). Regression coefficients are reported in Table S3; estimated subgroup changes are shown in Fig. 2.

We analyzed the first and corresponding authors separately because they often occupy distinct roles in the writing process [35]: First authors typically draft the manuscript, handling the initial text generation and subsequent revisions. This role directly confronts the linguistic challenge of producing fluent English prose. Corresponding authors, in contrast, generally oversee the project, focusing on conceptual framing and final editing for coherence. Comparing these two groups thus provides insight into how generative AI tools interact with different stages of the writing process and with varying degrees of linguistic demand and professional responsibility.

Among corresponding authors, publications associated with less-established researchers consistently showed larger post-ChatGPT increases in AI-assisted writing (Fig. 2). In both national linguistic contexts, increases were greater for authors outside the top 5% in cumulative productivity and citation impact, and for junior rather than senior authors. At the same time, publications associated with non-English-speaking countries generally showed larger increases within each author-profile group. The two dimensions together produced substantial overlap across groups: Increases in AI-assisted writing among more-established authors in non-English-speaking countries were similar to those among less-established authors in English-speaking countries. Together, these results indicate that heterogeneity in AI-assisted writing among corresponding authors is associated with both national linguistic context and researchers' professional and institutional status.

For first authors, differences associated with national linguistic context remained particularly pronounced. Across all comparisons, papers from non-English-speaking countries showed larger post-ChatGPT increases in AI-generated content than those from English-speaking countries (Fig. 2). However, differences between more established and less established scientists were evident in papers from English-speaking countries but were far smaller among those from non-English-speaking countries. Specifically, the observed difference in the change of AI-generated content between top and non-top authors, as measured by cumulative citations, was modest ($\Delta = -0.005$, $p < .001$). The difference was not statistically significant between highly productive authors and other authors ($\Delta = -0.002$, $p = .06$), nor was it significant between senior and junior authors ($\Delta = -0.002$, $p = .06$). This contrasts with the patterns of corresponding authors, where professional status is associated with the increase in AI-assisted writing in both English-speaking and non-English-speaking countries. For first authors, however, differences associated with national linguistic context were larger than those associated with productivity, citation impact, or seniority, suggesting broader diffusion of AI-assisted writing among researchers affiliated with non-English-speaking countries. This pattern may partly reflect differences in writing responsibilities: corresponding authors primarily refine and structure manuscripts, whereas first authors, especially those from non-English-speaking contexts, handle the initial drafting, where linguistic barriers are more pronounced, and AI tools offer greater utility [35]. However, we do not directly observe individual authors' writing responsibilities and therefore treat this interpretation cautiously.

In addition to the cumulative dimensions of academic status, we further examined whether AI adoption may also vary with more immediate publication demands. Specifically, we conducted another analysis based on recent publication velocity, defined as the number of an author's publications during the focal article's publication year and the immediately preceding year (Fig. S7). For corresponding authors, differences in AI-assisted writing between higher- and lower-velocity groups were small. Among first authors, however, those with higher recent publication velocity exhibited larger increases in estimated AI-assisted writing, particularly in non-English-speaking countries. Because publication velocity captures recent output intensity rather than publication pressure directly, we interpret this pattern as evidence that AI adoption also varies with authors' more immediate publication activity.

Finally, AI-assisted writing also varied by institutional prestige. Papers from top-ranked institutions exhibited smaller increases in AI-assisted writing than those from lower-ranked institutions, regardless of country (Fig. 2d). Moreover, the difference between papers from English-speaking, lower-ranked institutions and those from non-English-speaking, top-ranked institutions was statistically insignificant, suggesting that institutional prestige may be associated with the magnitude of adoption differences across national linguistic contexts. Results for first and corresponding authors were highly similar, as expected, since they are very likely to be associated with the same institution or institutions of similar standing. Taken together with the preceding analyses, these results show that heterogeneity in AI-assisted writing is associated not only with individual author characteristics but also with institutional context, although the specific mechanisms underlying this association are not directly observed in our data.

**Greater Use of AI-assisted Writing is Associated with Prior AI Research Experience and Changes in Productivity**

The preceding analyses, based on aggregations of publications, revealed population-level patterns but did not directly examine how individual scientists' behavior changed over time. To address this, we conducted an author-level analysis restricted to biomedical scientists who published in both the pre- (2021–2022) and post-ChatGPT (2023–2024) periods (Methods). This design enabled us to track changes in LLM usage by individual authors. Results from the DiD analysis with author fixed effects (Table S4) and the complementary author-level OLS regression (Table S5) are consistent with the differences across national linguistic contexts identified in our earlier analyses: scientists affiliated with non-English-speaking countries showed larger increases in AI-generated content post-ChatGPT than their peers affiliated with English-speaking countries.

We next examined whether prior exposure to AI research was associated with greater use of AI-assisted writing. Authors who had previously published in AI-related fields before 2023 showed an additional increase in the fraction of AI-generated sentences after the release of ChatGPT (first author: $\Delta = 0.013$, $p < .001$; corresponding author: $\Delta = 0.011$, $p < .001$) (Figs. 3a–b, Table S5). Moreover, the association between prior AI research experience and increased AI-assisted writing was stronger among researchers affiliated with non-English-speaking countries, with an additional increase in the estimated fraction of AI-generated content of 0.013 for first authors ($p < .001$) and 0.005 for corresponding authors ($p = .014$), relative to researchers affiliated with English-speaking countries. This interaction suggests that prior AI research experience is associated with larger increases in AI-assisted writing particularly in non-English-speaking national contexts.

We further examined whether greater use of AI-assisted writing is associated with changes in scientific productivity. For each scientist, we calculated the change in AI-generated content in papers and the change in number of publications between pre- and post-ChatGPT periods; the results reveal a modest but positive correlation (first author: $\beta = 2.456$, $p < .001$; corresponding author: $\beta = 3.334$, $p < .001$) (Figs. 3c–d, Table S6). Average productivity declined slightly compared to the 2021–2022 period, which may partly reflect the end of the pandemic's temporary boost to publication activity [36]. However, authors who exhibited larger increases in AI-assisted writing (top 40% by increase in AI-generated content) tended to publish slightly more papers after ChatGPT's release. Furthermore, this association was stronger among researchers from non-English-speaking countries, with the estimated association larger by 0.382 units for first authors ($p < .001$) and 0.400 units for corresponding authors ($p < .001$) relative to researchers affiliated with English-speaking countries (Table S6). The productivity gap between English- and non-English-speaking countries was most evident among those with minimal increases in AI use and narrowed as AI adoption rose. This pattern suggests that a greater increase in AI-assisted writing is associated with a modest narrowing of productivity differences between the two country groups.

## Discussion

This study examines the uneven diffusion of AI-assisted writing across national linguistic, professional and institutional contexts in global biomedical research. Scientific publications associated with non-English-speaking countries, particularly those with lower national English proficiency, showed greater post-ChatGPT increases in AI-assisted writing. Adoption also varied systematically across author profiles, with greater increases among less-published, less-cited, and earlier-career scientists and those at lower-ranked institutions. Prior AI research experience was associated with greater increases in AI-assisted writing, and greater increases in AI-assisted writing were modestly associated with greater increases in publication productivity. Collectively, these patterns show that AI-assisted writing is likely to be adopted unevenly across the global biomedical research community, with the largest increases occurring among groups that have historically been associated with greater barriers to scientific publishing.

AI-assisted writing represents a new writing technology and its adoption patterns can be interpreted in part through the diffusion-of-innovations framework developed by Rogers [15]. In this framework, adoption depends on the perceived relative advantage, compatibility, trialability, observability, and complexity of an innovation. Compared with traditional translation software or professional editing and proofreading services, LLMs provide faster and more affordable support for academic writing practice. Their simplicity and trialability further accelerate diffusion: they are intuitive, require minimal technical skill, and often offer free trials or low-cost access. The heterogeneity observed in our study may be particularly informative about differences in perceived relative advantage and compatibility across researcher contexts. In terms of relative advantage, AI-assisted writing may offer greater marginal benefits in contexts where researchers face greater linguistic or professional constraints. For researchers working in non-English-speaking environments, these advantages directly address a persistent constraint in scientific communication. Conversely, the smaller increases observed among scientists from English-speaking countries, more established researchers, and those at higher-ranked institutions could be consistent with smaller marginal benefits where writing routines or alternative forms of support are more established. Compatibility may also help interpret the greater increases observed among researchers with prior AI research experience, for whom LLM-based tools may fit more readily with existing technical knowledge and professional practices. Similarly, the smaller increases among senior researchers could be consistent with greater attachment to established writing routines or professional norms. They may also perceive reputational, authorship, or disclosure-related concerns associated with AI use in scientific writing. Thus, the observed heterogeneity is nevertheless consistent with a diffusion process in which the perceived benefits and compatibility of the same technology differ substantially across linguistic and professional groups. However, these mechanisms are not directly measured in our data, and we therefore present them as theoretically informed interpretations rather than causal explanations of adoption.

Our findings have broader implications for inequalities in scientific communication. Researchers who face language barriers in publishing in English encounter additional burdens in reading, writing, editing, and revising manuscripts, and these barriers can affect participation and visibility in scientific communication [18–23]. For them, LLM-based writing tools offer a low-cost and scalable form of linguistic support that may enhance clarity and readability while shortening writing time. Researchers also operate within a broader publish-or-perish culture [37–39] in which sustained publication output is strongly incentivized. Against this background, the larger increases in AI-assisted writing that we observe in lower-English-proficiency countries and among less-established researchers raise the possibility that these tools may have compensatory potential by providing accessible support for scientific publishing. Consistent with this possibility, our supplementary analyses show that first authors with higher recent publication velocity showed larger increases in AI-assisted writing, particularly in non-English-speaking countries (Fig. S7). Although publication velocity does not directly measure publication pressure, this pattern suggests that AI adoption may also vary with more immediate levels of research output intensity. Taken together, these findings are consistent with generative AI functioning as a potentially compensatory technology, with adoption increasing most rapidly among researchers who have historically faced greater linguistic, institutional, and career-related barriers.

However, reducing barriers to dominant scientific-English conventions is not necessarily equivalent to preserving linguistic diversity. Emerging evidence suggests that AI-generated texts rely heavily on Standard American English conventions and become more similar to U.S. scientific English [27–29]. Together, these findings point to a potential tension between reducing linguistic barriers and preserving linguistic diversity: GenAI may lower the costs of conforming to established scientific-English conventions while simultaneously narrowing variation in scientific expression. Our study does not directly test whether AI-assisted writing produces textual convergence, and future work should assess whether within-subfield stylistic diversity declines more rapidly in contexts and among researcher groups showing greater increases in AI-assisted writing.

Meanwhile, AI-assisted writing also raises broader concerns about incentives, research quality, and accountability. Excessive automation could flood editorial systems with low-quality or redundant submissions [40,41], as illustrated by the recent National Institutes of Health decision to limit individual investigators to six annual grant applications after reports of AI-generated proposals [42]. These challenges are amplified by the persistent publish-or-perish culture that prioritizes quantity over quality in research evaluation [43]. When career advancement depends heavily on publication counts, scholars, especially early-career researchers and those in competitive environments, may be more tempted to overuse generative tools to increase output. To safeguard research integrity, evaluation systems must therefore shift toward rewarding quality, reproducibility, and responsible practice rather than sheer productivity.

Yet, the use of AI in scientific research is likely inevitable, extending far beyond its use in academic writing. As AI and LLMs continue to evolve and become embedded in research and communication workflows [44–52], the central challenge is not whether to use AI but how to ensure its responsible integration. Addressing these risks will require coordinated action among publishers, funders, and research institutions. Clear and consistent disclosure standards are needed to distinguish between permissible language assistance and intellectual authorship, along with accountability mechanisms for verifying AI-assisted content. Ethical guidelines should also govern the use of sensitive or proprietary data in LLM training and application [53]. More broadly, responsible adoption will depend on a cultural shift, from viewing LLMs merely as tools for productivity to recognizing them as technologies that demand ethical oversight, transparency, and shared responsibility across the research ecosystem. Robust justifications for AI use, coupled with institutional guidance and researcher accountability, will be essential to ensure that technological progress strengthens rather than undermines the integrity, equity, and trustworthiness of global science [54].

Several limitations should be considered when interpreting our findings. First, our analyses are based on full text from PMC, which is not a random sample of global scientific publishing. Its composition is shaped by the availability of reusable full text, journal and publisher participation, and open- or public-access and deposition policies. Moreover, biomedical research differs from other fields in collaboration structures, manuscript organization, writing conventions, and the roles of argumentation, authorial voice, and language. These differences may shape both the opportunities for and use of AI-assisted writing. Our findings should therefore be interpreted as evidence from a large but selected biomedical corpus, rather than as estimates representative of the global scientific literature. Future research using multidisciplinary full-text corpora will be needed to assess the generalizability of these patterns. Our author- and institution-level measures also rely on OpenAlex metadata, which remain subject to missingness and author-disambiguation errors and may therefore introduce measurement error into longitudinal author profiles.

Second, our estimation of AI-assisted writing, while statistically validated, captures AI-associated linguistic patterns rather than all forms of AI assistance. Although validation analyses suggest that the estimator is relatively insensitive to light proofreading, it cannot determine the specific mode or degree of LLM involvement in an individual paper. It may also fail to capture uses of LLMs that do not leave a strong textual signature, such as assistance with idea development, organization, or other conceptual aspects of the research and writing process.

Third, our analyses are observational and should not be interpreted as establishing causal relationships. The observed differences across linguistic, professional, and institutional contexts may reflect multiple underlying mechanisms that are not directly measured in our data. Future research combining large-scale observational evidence with surveys, interviews, or experiments could more directly examine researchers' motivations, attitudes, constraints, and decisions surrounding AI-assisted writing.

Fourth, our primary language-related classification captures the national linguistic environment associated with an author's institutional affiliation rather than the author's individual native language or English proficiency. This distinction is particularly important in immigration-receiving English-speaking countries, where many researchers may be non-native English speakers and may themselves face language-related barriers. If such researchers are more likely to use AI-assisted writing, their inclusion in the English-speaking-country group would tend to raise the estimated level of AI-assisted writing in that group and thereby attenuate the observed difference between English- and non-English-speaking countries. However, because individual native language and English proficiency are not observed in our data, we cannot determine the magnitude of this attenuation. As a complementary robustness check, we inferred authors' nationality from their names and obtained similar cross-group patterns (Fig. S9), but name-inferred nationality is also not a direct measure of individual language background. Neither approach fully captures multilingual identities, international mobility, or the linguistic composition of collaborative teams.

Despite these limitations, this work presents timely, data-driven evidence of the uneven diffusion of a new technology across the biomedical scientific community. Future research should monitor these developments and examine their long-term implications for research quality, peer review, and global equity in scientific communication. Ultimately, understanding and guiding this transformation will be essential to ensuring that the integration of GenAI into science enhances, rather than undermines, the inclusiveness, rigor, and integrity of the research enterprise.

## Methods

### Data Sources

We utilized full-text biomedical publications from the PubMed Central Open Access Subset, focusing on works accepted between 2021 and 2024. We selected PMC because our analysis requires large-scale access to structured full-text data with identifiable manuscript sections. The relatively standardized organization of biomedical articles further facilitates section-level comparisons across a large corpus. In addition, PMC provides manuscript-history metadata, including received and accepted dates, allowing us to align publications more closely with the period in which the manuscripts were written relative to the release of ChatGPT.

These records were then linked to OpenAlex to obtain structured author and institutional metadata. We selected OpenAlex [55] not because its metadata are without limitations, but because it offers a combination of features particularly suited to our research design. Most importantly, it provides author and institution identifiers linked to longitudinal publication and citation histories, allowing us to construct researcher-level measures of cumulative productivity, citation impact, career age, institutional affiliation, and prior AI research experience at scale. Its geographical coverage is also important for our cross-country analyses: recent evaluations suggest that OpenAlex provides comparatively broad representation of research from lower- and middle-income countries relative to Web of Science and Scopus [56]. In addition, its openly available data and snapshots facilitate reproducibility of the linkage and longitudinal analyses. We nevertheless recognize that OpenAlex contains incomplete or inaccurate metadata in some fields and that author disambiguation remains imperfect [56–60]. We therefore use OpenAlex as a scalable source of longitudinal author and institutional metadata while explicitly accounting for these limitations in interpreting the resulting measures. We retained only English-written journal articles and reviews. Our primary analytical dataset includes 2,059,622 PMC full-text publications accepted between 2021 and 2024 that were successfully matched with OpenAlex for additional metadata information. Further details are provided in Supplementary Note 1.

### Estimating the fraction of AI-generated content in research papers

We adopted a distribution-based framework to quantify the extent of LLM use in academic writing following Liang *et al.* [32]. This approach involves modeling a focal publication as a mixture of human-written and AI-generated sentence distributions and estimating the mixing coefficient (i.e., the fraction of AI-generated sentences) through a maximum likelihood estimation. Compared with individual-text classifiers, which often yield high false-positive rates, this approach provides a statistically grounded estimate of AI involvement at scale. The approach has been validated and applied in previous studies [10,61–64]. Test cases demonstrated that the model's prediction error for AI-generated content was below 3.5% [32]. As an additional validation, our estimates were positively associated with the AI-generated probability scores from external AI detectors (Fig. S10). Furthermore, the estimator is substantially less sensitive to minor proofreading than to more extensive rewriting or generation [32]. Because the estimator captures AI-associated linguistic patterns, however, it cannot determine the specific mode or degree of LLM involvement in an individual paper; we therefore interpret it as a measure of AI-assisted writing rather than as evidence of a particular type of LLM use.

We applied this framework to estimate the fraction of AI-generated content in each paper, focusing on their *Introduction* and *Discussion* sections. To ensure consistency across papers, we first standardized the titles of all sections to align with the widely adopted *Introduction*, *Methods*, *Results*, and *Discussion* (IMRaD) structure in biomedical science publications [65,66]. We focused on the *Introduction* and *Discussion* because these sections involve substantial rhetorical and interpretive work, including positioning the study relative to prior literature, motivating its contribution, interpreting findings, and articulating their implications [67,68]. By contrast, *Methods* and *Results* are more procedurally and empirically constrained and showed substantially lower estimated AI-assisted writing in our section-level analysis (Fig. S11). As a robustness test, we repeated the cross-group analyses using the *Methods* and *Results* sections combined. Although the estimated increases in AI-assisted writing were substantially smaller in these sections, the relative differences across linguistic contexts and author profiles were broadly consistent with our primary findings (Fig. S8). We did not include abstracts in this robustness analysis because their short length provides insufficient text for stable paper-level estimation under our distribution-based framework. Using spaCy's small English language model, we then performed sentence segmentation and tokenization on the *Introduction* and *Discussion* texts, retaining only papers with at least 30 sentences across both sections to ensure a sufficient textual basis for stable estimation. For each paper, we estimated the fraction of AI-generated sentences within these two sections. Further details are provided in Supplementary Note 2.

**Measuring the National Linguistic Environment of Authors' Affiliation Countries**

We characterized the national linguistic environment associated with each author's institutional affiliation using three complementary indicators that together approximate national familiarity with English. First, English-speaking status was derived from the GeoDist database [69], classifying countries as English-speaking when at least 20% of the population speaks English (Table S1). Second, adult English-language proficiency scores were obtained from the 2024 English Proficiency Index [34] to quantify population-level command of English. Third, to capture structural relatedness between English and other languages, countries were grouped by linguistic distance into Germanic (excluding English-speaking countries, e.g., Germany and the Netherlands), other Indo-European (e.g., France, Italy, and Russia), and non–Indo-European (e.g., China and Japan). We use English-speaking status as the primary country-level classification and the EPI and linguistic distance as complementary measures to examine whether differences in AI-assisted writing are systematically associated with independent dimensions of countries' linguistic relationship to English. We analyzed the affiliation countries of first and corresponding authors separately, who generally represent the lead and senior contributors of each paper [35]. More details are provided in Supplementary Note 1.3.

Importantly, these measures characterize the linguistic environment of an author's affiliated country and should not be interpreted as measures of the author's native language or individual English proficiency. This distinction is particularly relevant given the international mobility of the scientific workforce. As an additional robustness check that does not rely on institutional affiliation, we inferred scientists' nationality from their names using the Python package name2nat [70], and assigned each scientist the nationality with the highest predicted probability. Repeating the cross-group analysis using this alternative classification yielded similar patterns (Fig. S9). Because name-inferred nationality also does not directly measure native language or English proficiency, we treat this analysis as an alternative nationality-based operationalization rather than as an individual-level measure of linguistic background.

**Regression Analysis**

*Difference-in-Differences Analysis of Variation in LLM Use*

We used a difference-in-differences framework to compare the changes in AI-generated content in publications associated with English-speaking countries and those with non-English-speaking countries after the release of ChatGPT in late 2022. The model is specified as follows:

$$
\begin{aligned}
LLM_use_{i,j,s,t} = \ & \beta 0 + \beta 1 \cdot EnglishSpeakingCountry_i + \beta 2 \cdot PostGPT_t \\
& + \beta 3 \cdot EnglishSpeakingCountry_i \cdot PostGPT_t + \boldsymbol{Control_i} \\
& + \alpha_j + \mu_s + \varepsilon_{i,j,s,t}
\end{aligned} \tag{1}
$$

Here, $LLM_use_{i,j,s,t}$ is the estimated fraction of AI-generated sentences for paper $i$ accepted in journal $j$ in subfield $s$ at time $t$, $EnglishSpeakingCountry_i$ is a binary variable indicating whether the first or corresponding author of paper $i$ is affiliated with an English-speaking country. $PostGPT_t$ is a binary variable indicating whether paper $i$ was accepted in the post-GPT period (2023–2024). The interaction term $\beta 3$ is the DiD estimator, capturing the additional change in AI-generated content among papers from English-speaking countries relative to those from non-English-speaking countries, under the parallel-trend assumption (Fig. S1). $\boldsymbol{Control_i}$ is a vector of control variables including the log-transformed author count, reference count, and sentence count in the examined text. We further included the journal fixed effect ($\alpha_j$) and subfield fixed effect ($\mu_s$) to control for unobserved heterogeneity across publication venues and subfields. $\varepsilon_{i,j,s,t}$ indicates the error term.

*Difference-in-Difference-in-Differences Analysis of Heterogeneity Across Author Profiles*

To examine whether the association varies across different author profiles, we extended the model to a difference-in-difference-in-differences framework. Specifically, we classified publications from English- and non-English-speaking countries into sub-groups based on four author characteristics, namely productivity, citation impact, career seniority, and institutional prestige.

For productivity and citation impact, we classified each paper based on whether the first or corresponding author ranked among the top 5% scientists within the same subfield and the same year (by cumulative publication and citation counts in OpenAlex); we used top 1%, top 10%, and top 20% as alternative thresholds for robustness checks (Figs. S3 and S4). Career seniority of an author was measured as the number of years between the current paper and the author's first publication recorded in OpenAlex; authors with a career age lower than 15 years were defined as junior, and those with a career age of 15 or more as senior; we used 10 and 20 years as alternative thresholds for robustness checks (Fig. S5). Institution prestige was determined using the 2025 QS World University Rankings in Life Sciences and Medicine, with the top 100 institutions in the ranking as top institutions; alternative thresholds were also considered for robustness checks (Fig. S6). Based on the classification, we expanded the previous DiD model to a DDD model as follows:

$$LLM_use_{i,j,s,t} = \beta 0 + \beta 1 \cdot EnglishSpeakingCountry_i + \beta 2 \cdot PostGPT_t + \beta 3 \cdot AuthorProfileGroup_i + \beta 4 \cdot PostGPT_t \cdot EnglishSpeakingCountry_i + \beta 5 \cdot PostGPT_t \cdot AuthorProfileGroup_i + \beta 6 \cdot EnglishSpeakingCountry_i \cdot AuthorProfileGroup_i + \beta 7 \cdot PostGPT_t \cdot EnglishSpeakingCountry_i \cdot AuthorProfileGroup_i + \boldsymbol{Control_i} + \alpha_j + \mu_s + \varepsilon_{i,j,s,t} \quad (2)$$

Here, $AuthorProfileGroup_i$ is a binary indicator denoting whether paper $i$ falls into the higher author-profile category (e.g., papers authored by a senior scholar, a top-ranked scholar by productivity or citation, or a scholar affiliated with top 100 institutions). The triple interaction $\beta 7$ captures the DDD effect, which captures whether the DiD estimate comparing English- and non-English-speaking countries differs between the higher and lower author-profile groups. In our analysis, we used $\beta 2$, $\beta 4$, $\beta 5$, and $\beta 7$ to estimate the change in the fraction of AI-generated content for the four groups based on country of origin and author profile. For example, $AuthorProfileGroup_i$ = 1 indicates that paper $i$ is authored by a senior scientist, and 0 otherwise, then the average change for papers by senior and junior authors from an English-speaking country can be estimated through $\beta 2$ + $\beta 4$ + $\beta 5$ + $\beta 7$ and $\beta 2$ + $\beta 4$, respectively, while the average change for papers by senior and junior authors from a non-English-speaking country can be estimated through $\beta 2$ + $\beta 5$ and $\beta 2$, respectively. $\boldsymbol{Control_i}$ is a vector of control variables including the log-transformed author count, reference count, and sentence count in the examined text. $\alpha_j$ denotes journal fixed effect, $\mu_s$ denotes paper subfield fixed effect and $\varepsilon_{i,j,s,t}$ indicates the error term.

*Author-Level Analysis of AI-Assisted Writing Before and After ChatGPT*

To examine how individual authors' use of AI-assisted writing changed following the release of ChatGPT, we conducted an author-level difference analysis. We focused on scientists who published as first or corresponding authors in both pre-ChatGPT (2021–2022) and post-ChatGPT (2023–2024) periods. Among those active in 2021–2022, 198,629 (25.1%) first authors and 233,520 (40.7%) corresponding authors also published in 2023–2024. Authors who changed their affiliated countries between the two periods were excluded. Because many authors published only a small number of papers in each period, we used the maximum estimated fraction of AI-generated sentences across an author's papers within each period. This measure captures the strongest observed signal of AI-assisted writing for each author-period, rather than the author's average use.

We first identified potential author-level characteristics that are likely associated with changes in LLM use between the two periods. Our baseline OLS model is as follows:

$$\Delta LLM_use_{k,s} = \beta 0 + \beta 1 \cdot EnglishSpeakingCountry_k + \beta 2 \cdot \log(CumulativePubs_k) + \beta 3 \cdot \log(CumulativeCits_k) + \beta 4 \cdot CareerAge_k + \beta 5 \cdot TopInstitution_k + \mu_s + \varepsilon_{k,s} \quad (3)$$

Here $\Delta LLM_use_{k,s}$ represents the within-author change in the fraction of AI-generated content in papers from the 2021–2022 to the 2023–2024 period. $EnglishSpeakingCountry_k$ indicates whether author $k$ is affiliated with an English-speaking country. Control variables include the log-transformed cumulative numbers of publications and citations in 2022, and career age (years from first publication until 2022), as well as a dummy for affiliation in one or more top 100 institution(s) in 2022. Subfield fixed effects $\mu_s$ account for disciplinary differences, where an author's primary subfield is defined as the one in which they have the greatest number of publications. $\varepsilon_{k,s}$ indicates the error term.

We next examined whether one's prior experience in AI research moderated changes in AI-assisted writing. An indicator variable, $AI_author_k$, was introduced to identify authors who had published at least one paper in the subfield of Artificial Intelligence (as recorded by OpenAlex) before 2023. The model is expressed as follows:

$$\Delta LLM_use_{k,s} = \beta 0 + \beta 1 \cdot EnglishSpeakingCountry_k + \beta 2 \cdot AI_author_k + \beta 3 \cdot EnglishSpeakingCountry_k \cdot AI_author_k + \beta 4 \cdot \log(CumulativePubs_k) + \beta 5 \cdot \log(CumulativeCits_k) + \beta 6 \cdot CareerAge_k + \beta 7 \cdot TopInstitution_k + \mu_s + \varepsilon_{k,s} \quad (4)$$

The interaction term $\beta 3$ captures whether the association between prior AI research experience and changes in AI-assisted writing differs between authors from English- and non-English-speaking countries. $\mu_s$ indicates author subfield fixed effect. $\varepsilon_{k,s}$ indicates the error term.

Finally, we examined whether increased AI-assisted writing was associated with changes in an author's publication productivity. The model is specified as follows:

$$\Delta Pub_{k,s} = \beta 0 + \beta 1 \cdot EnglishSpeakingCountry_k + \beta 2 \cdot \Delta LLM_use_k + \beta 3 \cdot EnglishSpeakingCountry_k \cdot \Delta LLM_use_k + \beta 4 \cdot \log(CumulativePubs_k) + \beta 5 \cdot \log(CumulativeCits_k) + \beta 6 \cdot CareerAge_k + \beta 7 \cdot TopInstitution_k + \mu_s + \varepsilon_{k,s} \quad (5)$$

where $\Delta Pub_{k,s}$ represents the change in the number of publications by author $k$ between the pre- and post-ChatGPT periods. $\beta 3$ estimates whether the relationship between changes in LLM use and productivity differs between authors from English- and non-English-speaking countries. $\mu_s$ indicates author subfield fixed effect. $\varepsilon_{k,s}$ indicates the error term.

## Data Availability

The two main datasets used in this study are publicly available. The PubMed Central Open Access Subset is available at https://pmc.ncbi.nlm.nih.gov/tools/openftlist. The OpenAlex snapshot is available at https://help.openalex.org/access/snapshot/.

## Code Availability

The code can be accessed via GitHub at https://github.com/MetascienceLab/aiwriting [71].

## Acknowledgments

The authors received no specific funding for this work. The authors would like to thank three anonymous reviewers, all members of the Metascience Research Lab (MSRL) at University of Wisconsin-Madison, and all members of the Knowledge Discovery (KD) lab at Peking University for their insightful comments. This work was presented in many seminars and workshops. The authors acknowledge Zhichao Fang, Richard Freeman, Junzhi Jia, Ming Ren, Mike Thelwall, and all other audiences for their constructive comments. GPT-5.5 was used solely to assist with English proofreading and language refinement. The authors have carefully reviewed the final manuscript and take full responsibility for the content.

## Author Contribution Statement

J.L., Y.B., and C.N. conceived the study and designed the analyses. J.L., Y.H., and Z.Z. collected the data and performed the analysis. J.L. wrote the manuscript, and Y.B. and C.N. edited the manuscript.

## Competing Interests Statement

The authors declare no competing interests.

# Figures

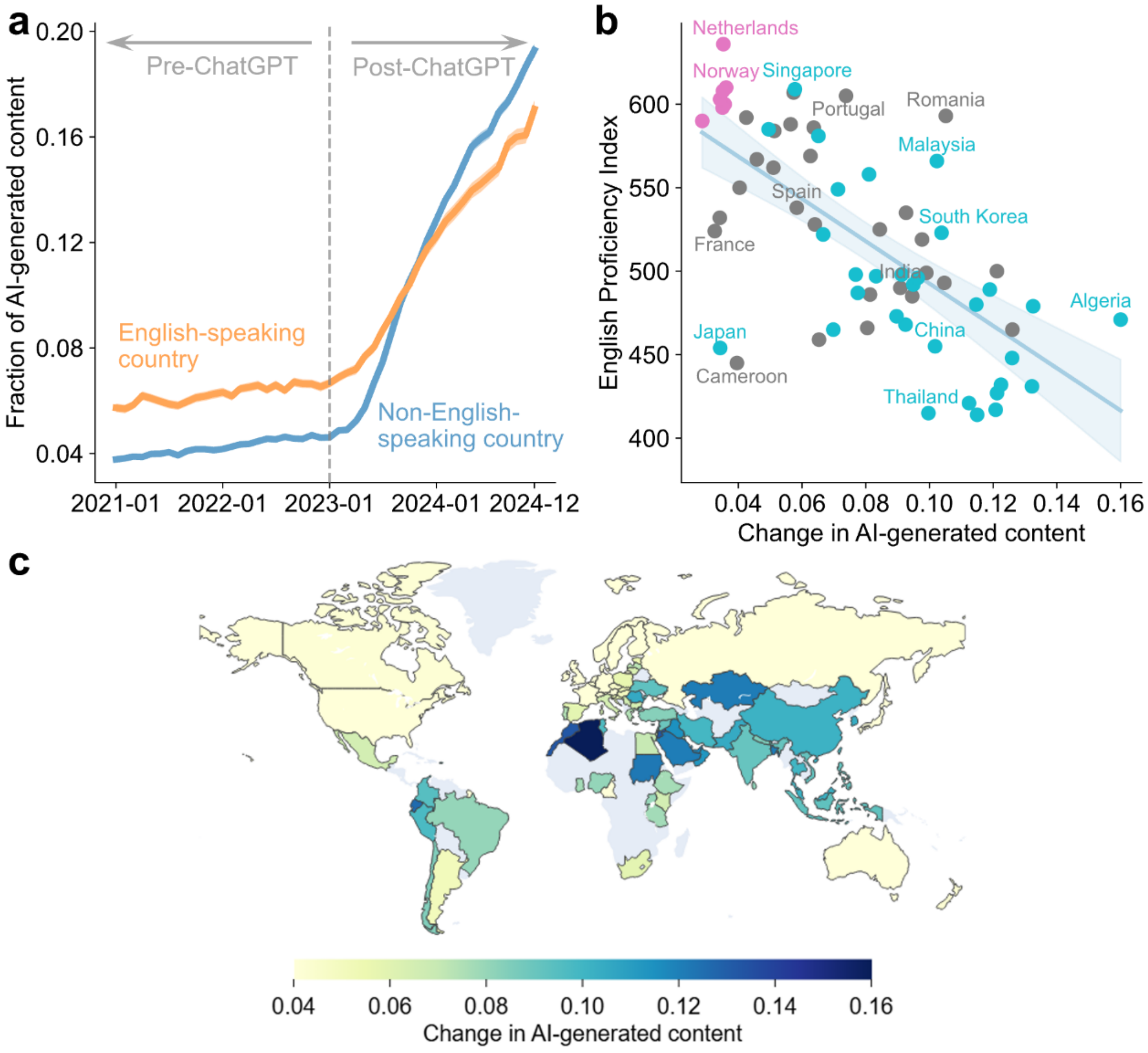

**Figure 1. Differential changes in AI-generated content across countries, based on affiliated countries of first authors. (a)** Monthly trends in the fraction of AI-generated content in papers from English- and non-English-speaking countries between 2021 and 2024. **(b)** Association between national English Proficiency Index scores and their changes in the fractions of AI-generated content. Countries where English is the only official language or without an EPI score were excluded. Countries are further grouped by linguistic distance to English: Germanic excluding English (pink), other Indo-European (gray), and non-Indo-European languages (cyan). **(c)** Country-level changes in the average fraction of AI-generated content in papers accepted during pre- and post-ChatGPT periods. Countries with fewer than 100 papers each year between 2021 and 2024 were excluded from this analysis and shown in grey. Across all panels, a

publication’s country assignment is based on its first author’s institutional affiliation; results are robust when using corresponding authors’ institutional affiliation (Fig. S2). Shaded regions denote the 95% confidence intervals.

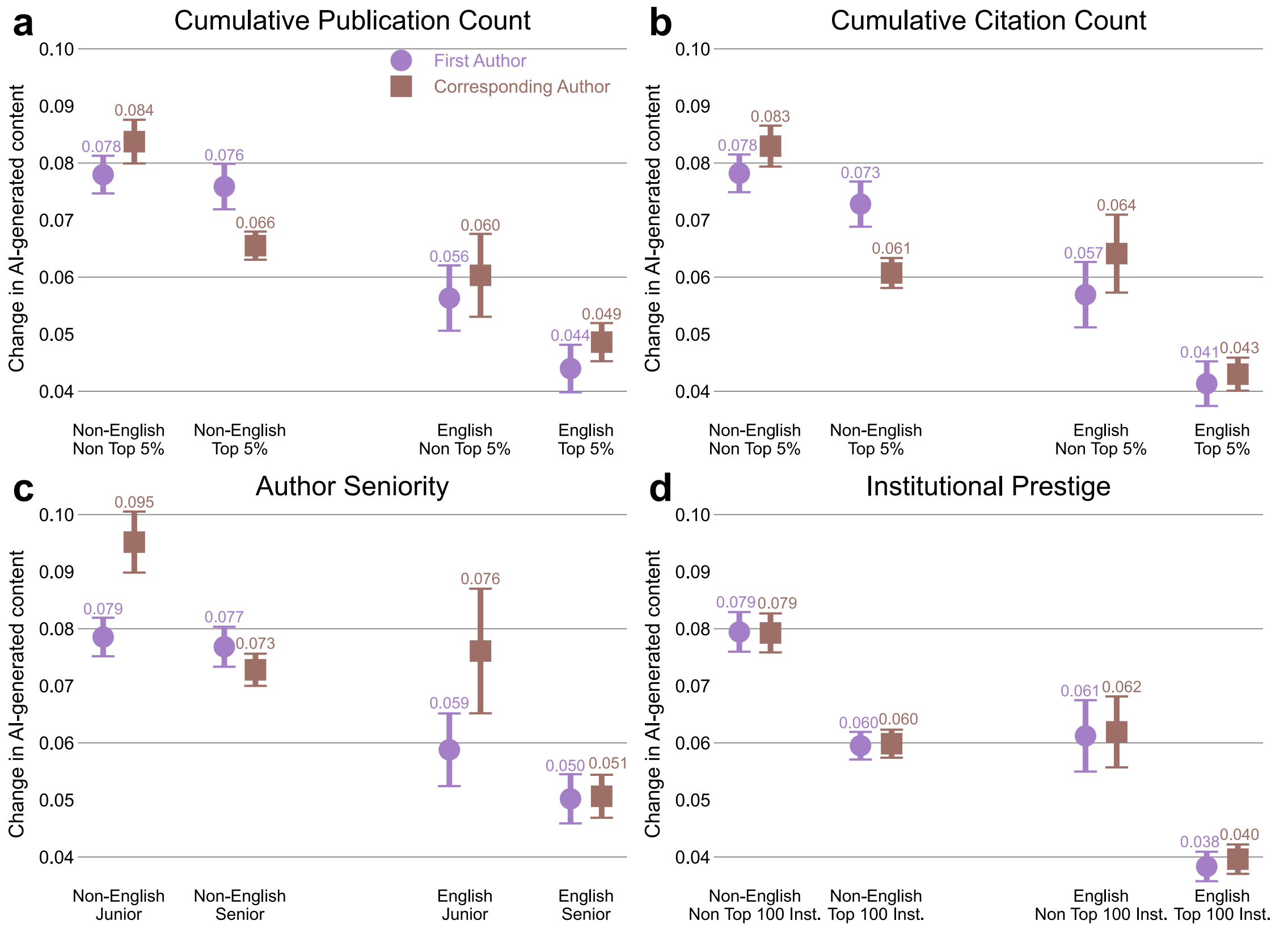

**Figure 2. Change in AI-generated content between pre- and post-ChatGPT periods, stratified by author profile and affiliation country group.** A publication was classified by whether the focal author was affiliated with an English- or non-English-speaking country, and by four author characteristics: **(a)** cumulative number of publications (top 5% vs. non-top 5%), **(b)** cumulative number of citations (top 5% vs. non-top 5%), **(c)** career seniority (senior vs. junior), and **(d)** institutional prestige (top 100 institutions vs. non-top 100). Values represent the change in the fraction of AI-generated content between pre- and post-ChatGPT periods, with 95% confidence intervals. Changes were estimated based on the difference-in-difference-in-differences model (Methods). Full regression estimates are reported in Table S3. Robustness checks using different thresholds for grouping authors based on their characteristics are shown in Figs. S3–S6. A robustness check estimating the AI-generated content in *Methods* and *Results* sections instead of *Introduction* and *Discussion* sections is shown in Fig. S8. A robustness check using name-inferred nationality as an alternative to affiliation-based country is shown in Fig. S9.

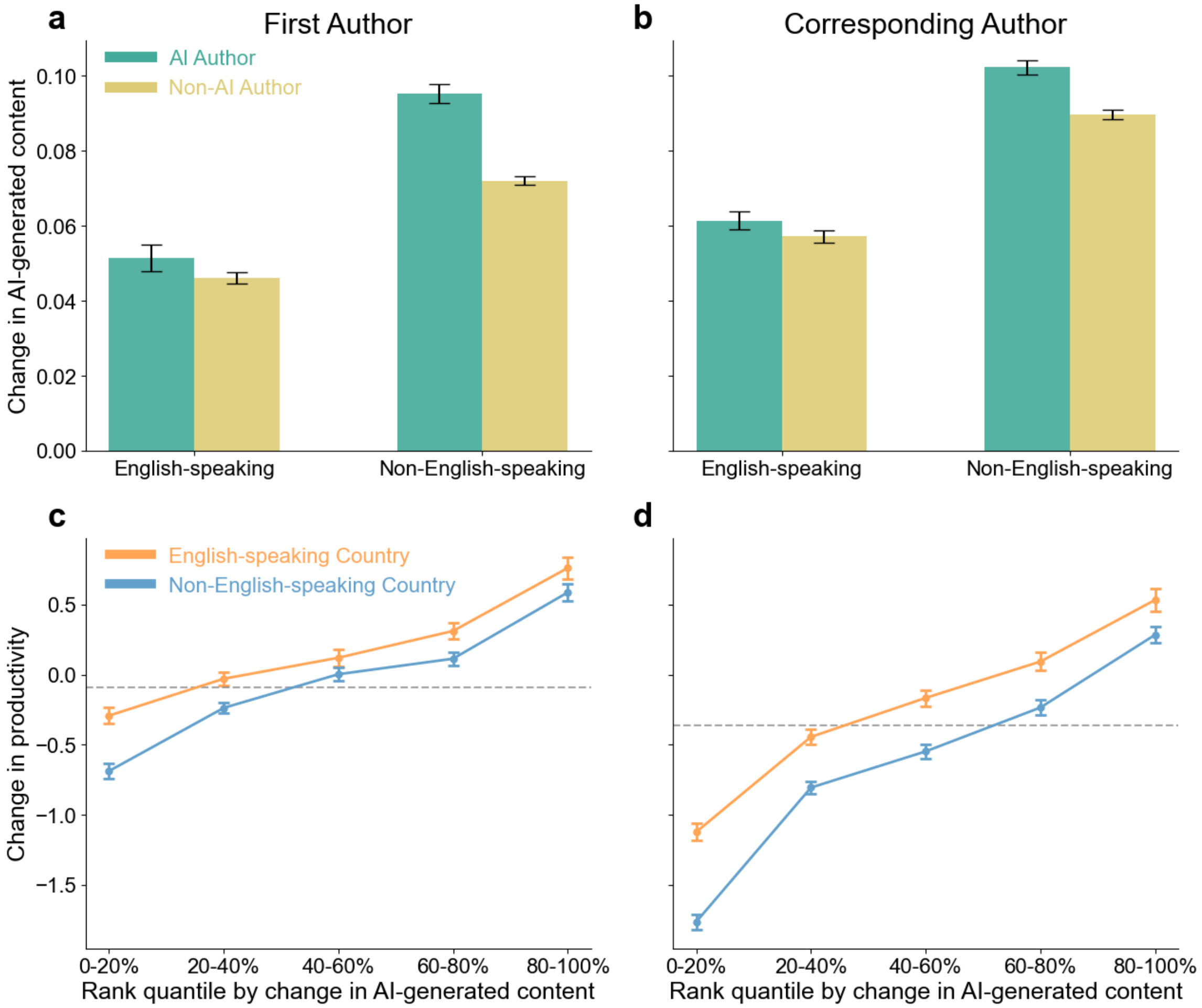

**Figure 3. Associations between authors' changes in AI-generated content, prior AI experience, and changes in publication productivity. (a)** Changes in the fraction of AI-generated content between pre- and post-ChatGPT periods based on first author's prior experience with AI. **(b)** Changes in the fraction of AI-generated content between pre- and post-ChatGPT periods based on corresponding author's prior experience with AI. **(c)** The relationship between change in LLM use and change in productivity between pre- and post-ChatGPT periods for first authors. **(d)** The relationship between change in LLM use and change in productivity between pre- and post-ChatGPT periods for corresponding authors. Scientists were assigned to five groups based on their rank quantile by the change in LLM use. The larger the rank quantile, the greater the increase in LLM use. The dashed line denotes the average change in productivity for all examined scientists. Error bars represent 95% confidence intervals. Regression estimates are reported in Tables S5 and S6.

# Supplementary Information for

## AI-Assisted Writing is Growing Fastest among Less Established Scientists in Non-English-Speaking Countries

**Authors:** Jialin Liu[1], Yongyuan He[2], Zhihan Zheng[3], Yi Bu[2, *], Chaoqun Ni[1, *]

[1] Information School, University of Wisconsin-Madison, Madison, WI 53706, U.S.A.

[2] Department of Information Management, Peking University, Beijing 100871, China

[3] Faculty of Education, The University of Hong Kong, Hong Kong 999077, China

* Corresponding Authors. Email: buyi@pku.edu.cn and chaoqun.ni@wisc.edu.

## Supplementary Note 1: Data sources and data preprocessing

### Supplementary Note 1.1: PubMed Central Open Access Subset

We obtained the main text of biomedical publications from PubMed Central (PMC) Open Access Subset [1] on May 21, 2026. PMC was selected because the current study requires large-scale, machine-readable full-text data with identifiable manuscript sections. Its structured XML files allow us to systematically extract and standardize comparable sections across publications, while the relatively standardized *Introduction*, *Methods*, *Results*, and *Discussion* (IMRaD) organization of biomedical articles facilitates section-level analysis at scale. PMC also provides manuscript-history metadata, including received and accepted dates, which more closely approximate the manuscript-writing timeline than publication dates alone. We parsed the XML files and extracted DOI, PubMed ID, PubMed Central ID, accepted dates, and main text. We retained papers accepted between 2021 and 2024. Given ChatGPT's release on November 30, 2022, and typical technology-adoption lags, we treat 2023–2024 as the post-ChatGPT period and 2021–2022 as the pre-ChatGPT period.

Regarding the main text, we mapped sections to the widely adopted IMRaD structure in biomedical scientific publications [2,3] and standardized the section titles. Specifically, we employed keyword-based identification using the following criteria: *Introduction* sections were identified using keywords "background", "intro", and "introduction"; *Methods* sections used "materials", "method", and "experiment" while excluding any sections containing "supplement" to avoid supplementary materials; *Results* sections were identified by the keyword "result"; and *Discussion* sections used "discussion" and "conclusion". We excluded sections belonging to multiple IMRaD standardized sections (e.g., "Results and Discussion"), as they might contain mixed content that would confound our analysis. We focused on the *Introduction* and *Discussion* because these sections involve substantial rhetorical and interpretive work, including positioning the study relative to prior literature, motivating its contribution, interpreting findings, and articulating their implications [4,5]. By contrast, *Methods* and *Results* are more procedurally and empirically constrained and showed substantially lower estimated AI-assisted writing in our section-level analysis (Fig. S11). As a robustness test, we repeated the cross-group analyses using the *Methods* and *Results* sections combined. Although the estimated increases in AI-assisted writing were substantially smaller in these sections, the relative differences across linguistic contexts and author profiles were broadly consistent with those observed in the *Introduction* and *Discussion* sections (Fig. S8). We retained only papers with at least 30 sentences in total across the *Introduction* and *Discussion* to ensure sufficient text for stable estimates.

**Supplementary Note 1.2: OpenAlex**

We linked the PubMed Central data to OpenAlex via their DOIs. OpenAlex [6] is a large-scale, open-access bibliographic dataset that also facilitates reproducibility. Recent studies have documented its broad coverage, while also noting continuing improvements and remaining deficiencies in specific metadata fields [7–11]. OpenAlex contains metadata on scholarly entities, including works, authors, sources, institutions, topics, publishers, and funders, as well as the connections between these entities. We used the publicly available OpenAlex snapshot dated March 30, 2026. Of the focal PMC papers, 98.2% were successfully matched to valid OpenAlex work IDs. We retained only English-written journal articles and reviews. Our final dataset includes metadata for 2,059,622 PMC full-text publications accepted between 2021 and 2024.

OpenAlex was selected because it provides structured author and institution identifiers linked to longitudinal publication and citation records, which allowed us to construct the author-level and institutional measures used in this study. In addition, geographical coverage is particularly important for our cross-country analysis. Maddi *et al.* compared OpenAlex with Scopus and Web of Science and found that OpenAlex provides comparatively greater representation of lower-middle-income and low-income countries [8].

We nevertheless recognize limitations in OpenAlex metadata. For example, institutional affiliation information can be incomplete, and previous studies have documented substantial variation in affiliation coverage across records and indexing sources [12,13]. In the PMC-linked sample used in this study, institutional information was missing for 2.8% of first authors and 3.9% of corresponding authors, indicating substantially higher affiliation completeness than has been reported for OpenAlex as a whole.

OpenAlex also performs author entity resolution by combining multiple bibliographic signals, including author names, coauthors, institutional affiliations, research topics, citation patterns, and ORCID identifiers when available. As with all large-scale bibliographic disambiguation systems, splitting and merging errors remain possible. We therefore use OpenAlex Author IDs as reproducible operational identifiers for constructing longitudinal author profiles rather than treating them as error-free identifiers of individual researchers.

**Supplementary Note 1.3: Measuring National Linguistic Environment of the Author Affiliation Countries**

To characterize the national linguistic environment associated with an author's institutional location, we used OpenAlex to identify each author's country or region of institutional affiliation (hereafter, country for simplicity). We analyzed the affiliation countries of first and corresponding authors separately, who generally represent the lead and senior contributors of each paper. Following the common convention in biomedical sciences, we defined the corresponding author as the last author.

Our primary country-level classification distinguished English- and non-English-speaking countries. We used the GeoDist dataset [14], which identifies languages spoken by at least 20% of a country's population. We classified an English-speaking country as one where English meets this 20% threshold (Table S1), including countries where it is a common or co-official language (e.g., Canada, Singapore). We excluded publications where the focal author was affiliated with both an English-speaking and a non-English-speaking country (2.2% of first authors; 2.1% of corresponding authors).

We further incorporated two complementary country-level linguistic indicators. First, we used scores from the 2024 English Proficiency Index (EPI) [15], which measures countries by the English-language proficiency of adults who took the EF Standard English Test. The EPI inherently excludes many primary English-speaking countries like the U.S. and Canada, assuming high proficiency. Second, to capture structural relatedness between English and other languages, we grouped countries according to the linguistic distance of their official languages from English: Germanic (e.g., Germany and the Netherlands), other Indo-European (e.g., Russia and France), and non-Indo-European (e.g., China and Japan). Together with the binary English-speaking-country classification, these measures allow us to examine whether differences in AI-assisted writing are systematically associated with multiple dimensions of countries' linguistic relationship to English.

Importantly, these measures characterize the linguistic environment of an author's affiliated country and should not be interpreted as measures of the author's native language or individual English proficiency. This distinction is particularly relevant given the international mobility of the scientific workforce. As an additional robustness check that does not rely on current institutional affiliation, we inferred scientists' nationality from their names using the Python package name2nat [16], and assigned each scientist the nationality with the highest predicted probability. Repeating the cross-group analysis using this alternative classification yielded similar patterns (Fig. S9). Because name-inferred nationality also does not directly measure native language or English proficiency, we treat this analysis as an alternative nationality-based operationalization rather than as an individual-level measure of linguistic background.

## Supplementary Note 2: Estimating the fraction of AI-generated sentences in scientific publications

### Supplementary Note 2.1: A brief review

Existing studies on estimating AI-generated content in academic writing have primarily employed a keyword-based approach. This technique usually identifies abrupt increases in the frequency of specific "style words" linked to large language models (LLMs). For example, Stokel-Walker noticed that the term "commendable" appeared 240 times in Scopus in 2020 and rose to 829 in 2023 (a 245% increase) [17]. Similarly, OpenAlex data showed that the frequency of "delve" rose from 0.056% in 2022 to 0.793% in 2024 [18]. Following a similar approach, Kobak *et al.* [19] studied 15 million biomedical abstracts and estimated that at least 13.5% of 2024 abstracts were processed by LLMs. While straightforward, these keyword-based approaches are criticized for being arbitrary, as they do not account for the text's broader semantic information.

Other studies have examined AI-generated content through AI detectors. Unlike keyword-based approaches, AI detectors consider whole-document-level or sentence-level features to provide a more specific metric. Various detection tools were adopted in previous studies, such as Originality.ai [20], GPTZero [21], and Smodin [22]. However, the reliability and interpretability of AI detection tools remain questionable [23,24]. For instance, an analysis of 12 publicly available tools and two commercial systems revealed overall accuracy rates below 80% [24]. These tools are prone to both false positives and false negatives, and their accuracy degrades significantly when text is edited by humans or rewritten by other machines. OpenAI [25] has echoed this concern, stating that its own research found detectors unreliable: "Our research into detectors didn't show them to be reliable. While other developers have released detection tools, we cannot comment on their utility."

Liang *et al.* [26] employed a distinct, population-level framework to estimate LLM-modified content. This method enables large-scale quantitative analysis by comparing the word frequency distribution of the target document against known LLM-modified and human-written corpora. Applying this framework to over one million papers from arXiv, bioRxiv, and Nature portfolio journals, they identified a steady increase in LLM usage in *Abstract* and *Introduction*.

### Supplementary Note 2.2: Estimation approach adopted by this study

We followed the population-level framework proposed by Liang *et al.* [26] to quantify the extent of LLM usage in academic writing by analyzing word distribution. The approach involves modeling a focal paper as a mixture of human-written and AI-generated sentence distributions and estimating the mixing coefficient (i.e., the fraction of AI-generated sentences) through maximum likelihood estimation (MLE). We adopted the following step-by-step procedure for estimation:

*Word occurrence probabilities estimation*

For each word $t$, let $\hat{p}_t$ and $\hat{q}_t$ denote its occurrence probabilities in human-written and AI-generated sentences, respectively:

$$\hat{p}_t = \frac{Number\ of\ human\ sentences\ containing\ t}{Total\ number\ of\ human\ sentences} \quad (1.)$$

$$\hat{q}_t = \frac{Number\ of\ AI\ sentences\ containing\ t}{Total\ number\ of\ AI\ sentences} \quad (2.)$$

As we focused on biomedical publications from PMC in the current study, we utilized the occurrence probabilities estimated from bioRxiv in previous work [26] due to the clear disciplinary overlap.

*Sentence likelihood modeling*

The likelihood of a sentence $x_i$ under the human, $P(x_i)$, and AI, $Q(x_i)$, is defined as:

$$P(x_i) = \prod_{t \in x_i} \hat{p}_t \times \prod_{t \notin x_i} (1 - \hat{p}_t) \quad (3.)$$

$$Q(x_i) = \prod_{t \in x_i} \hat{q}_t \times \prod_{t \notin x_i} \left(1 - \hat{q}_t\right) \quad (4.)$$

*Maximum likelihood estimation (MLE)*

For each sentence $x_i$, the mixture likelihood is calculated as:

$$L_i(\alpha) = (1 - \alpha)P(x_i) + \alpha Q(x_i) \quad (5.)$$

where $\alpha$ is the fraction of AI-generated sentences in the document and parameterizes the sentence-level mixture likelihood.

For $n$ sentences in the document, the log-likelihood $\mathcal{L}(\alpha)$ is given by:

$$\mathcal{L}(\alpha) = \sum_{i=1}^{n} log[L_i(\alpha)] \quad (6.)$$

$\alpha$ is estimated by maximizing the log-likelihood under the mixture distribution:

$$\hat{\alpha} = \arg \max_{\alpha \in [0,1]} \mathcal{L}(\alpha) \quad (7.)$$

**Supplementary Note 2.3: The validation of the utilized strategy**

The methodology used in this study has been evaluated through several validation exercises. First, the approach was previously validated and applied in different studies [27–31]. Liang *et al.* (2025) benchmarked this approach under multiple known mixture settings and reported population-level prediction errors below 3.5%. This approach was also evaluated under minor text edits. Their tests show that light LLM proofreading increased the estimated fraction by approximately one percentage point relative to the baseline. By contrast, restructuring produced estimates of approximately 50%, while fully AI-generated text produced estimates as high as 97%. This result indicates that the estimator is less sensitive to proofreading than extensive rewriting or generation. Second, we conducted an additional validation for the current study. We compared our results against our previous analysis [22] that used online detectors to generate AI probability scores (0 to 1) for over 5,000 abstracts. We grouped these abstracts into probability bins (e.g., [0-0.2), [0.2-0.4)) and applied our corpus-level estimation to each. As shown in Fig. S10, our estimates are positively correlated with the average AI-generated probability based on detection tools, showing correspondence with an external detector-based measure.

## Supplementary Figures

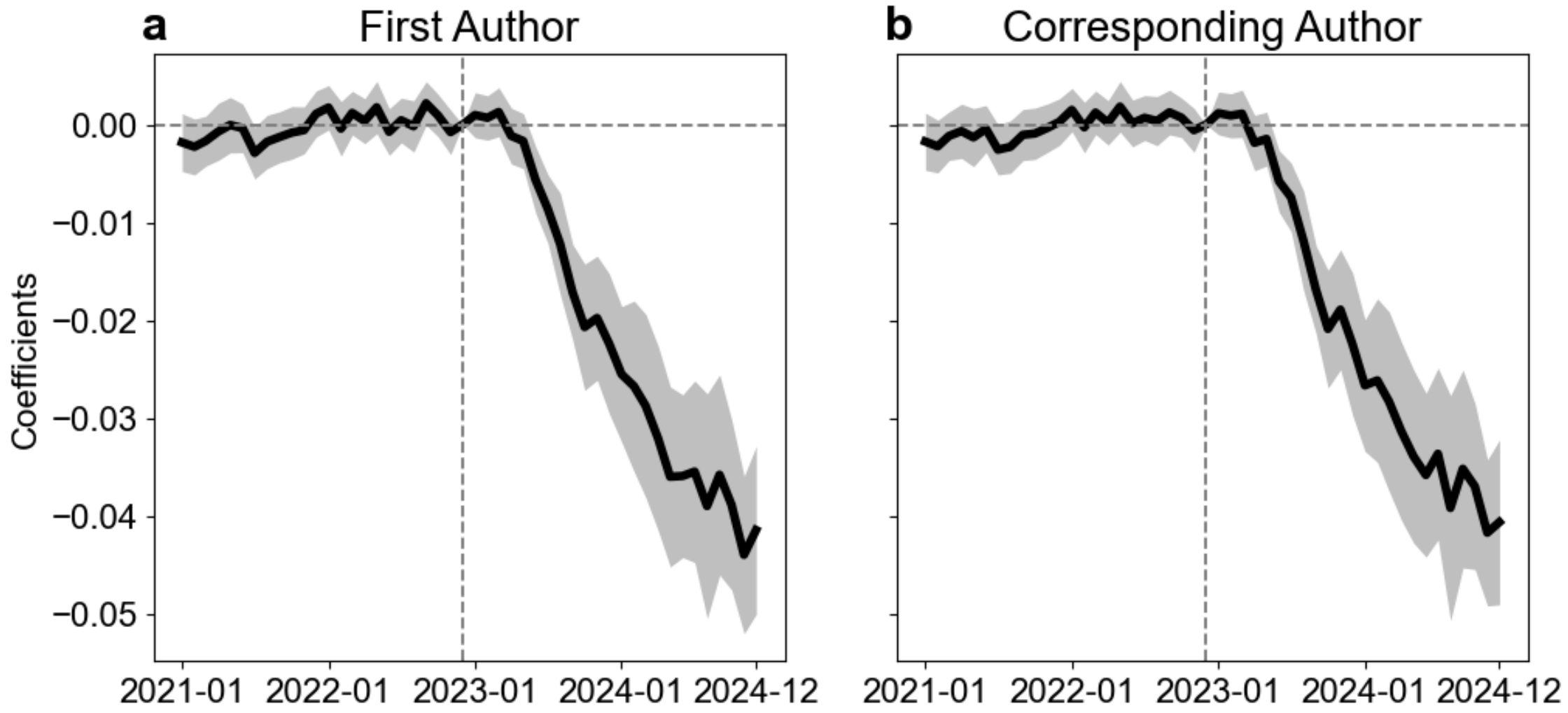

**Figure S1. Validation of the parallel trend assumption for the difference-in-differences analysis.** The plots show event study coefficients based on the country affiliations of **(a)** first authors and **(b)** corresponding authors. Each coefficient represents the estimated monthly difference in the fraction of AI-generated content between English- and non-English-speaking countries, relative to the December 2022 baseline. Shaded areas represent 95% confidence intervals. During the pre-ChatGPT period (2021–2022), the estimated coefficients are generally small and show no systematic temporal pattern, with the confidence intervals largely overlapping zero. This indicates that both groups exhibited parallel trends in AI-generated content before the event, consistent with the parallel-trends assumption.

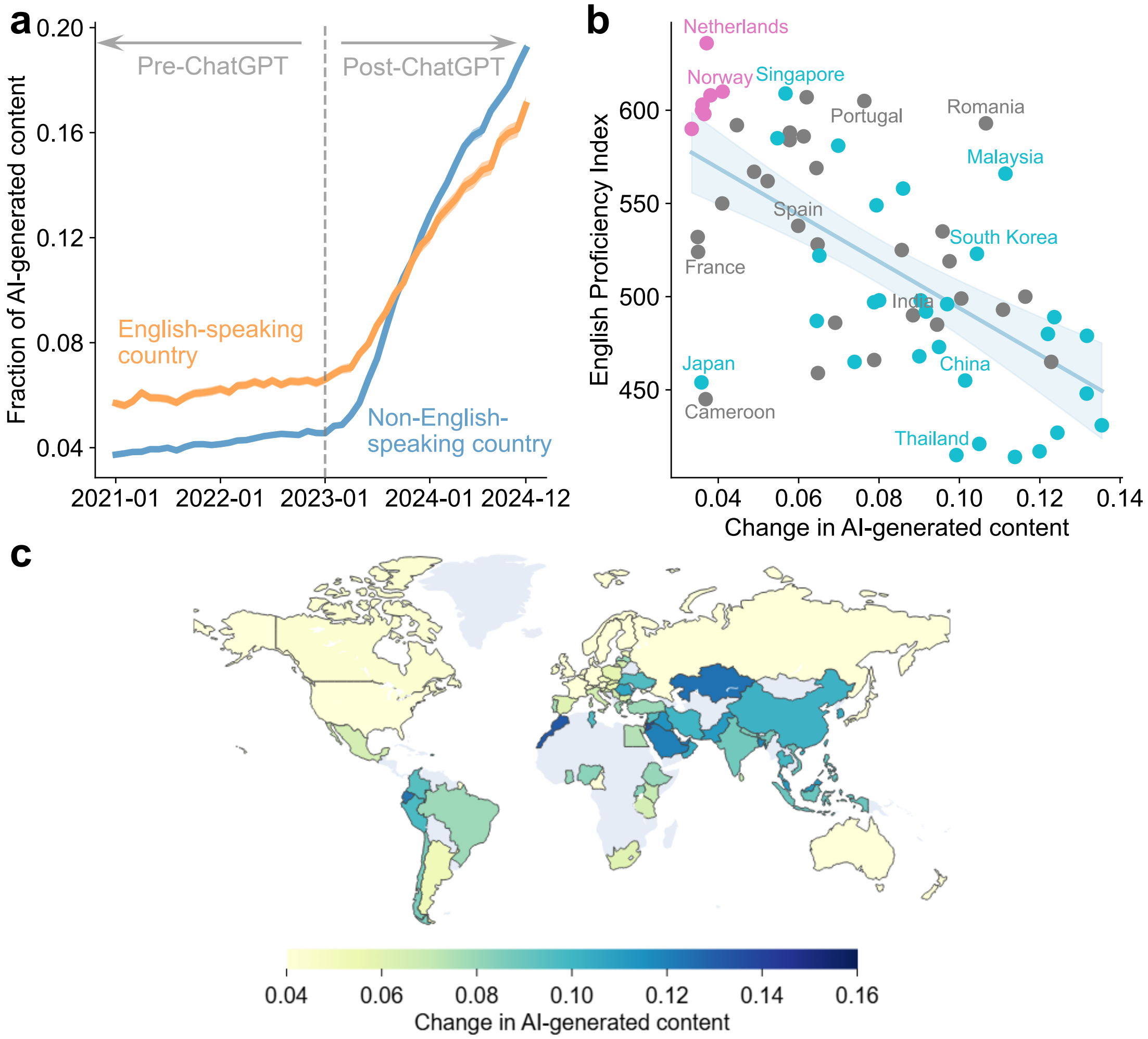

**Figure S2. Differential changes in AI-generated content across countries, based on affiliated countries of corresponding authors. (a)** Monthly trends in the fraction of AI-generated content in papers from English- and non-English-speaking countries between 2021 and 2024. **(b)** Association between national English Proficiency Index scores and their changes in the fractions of AI-generated content. Countries where English is the only official language or without an EPI score were excluded. Countries are further grouped by linguistic distance to English: Germanic excluding English (pink), other Indo-European (gray), and non-Indo-European languages (cyan). **(c)** Country-level changes in the average fraction of AI-generated content in papers accepted during pre- and post-ChatGPT periods. Countries with fewer than 100 papers each year between 2021 and 2024 were excluded from this analysis and shown in grey. Shaded regions denote the 95% confidence intervals.

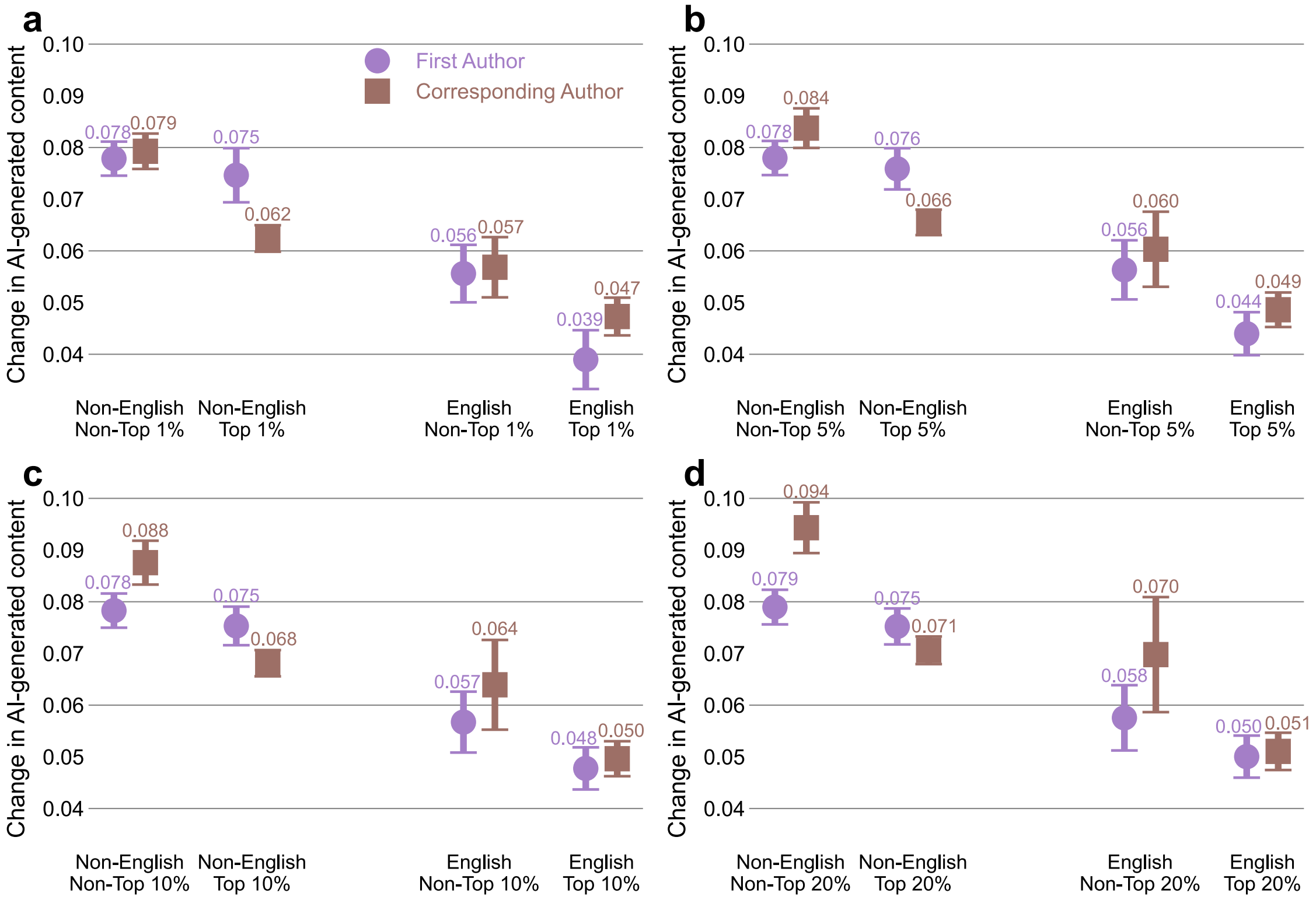

**Figure S3. Robustness check for the change in AI-generated content, stratifying authors by varying productivity thresholds.** The figure shows the mean change in AI-generated content between pre- and post-ChatGPT periods, comparing authors from English- and non-English-speaking countries. Authors are designated as "top" or "non-top" based on their cumulative publication count. To test the robustness of this comparison, the analysis is repeated using four different thresholds to define the "top" group: **(a)** top 1%, **(b)** top 5%, **(c)** top 10%, and **(d)** top 20%. Results are shown separately for first authors (purple circles) and corresponding authors (brown squares). Values represent the change in the fraction of AI-generated content between pre- and post-ChatGPT periods, with 95% confidence intervals.

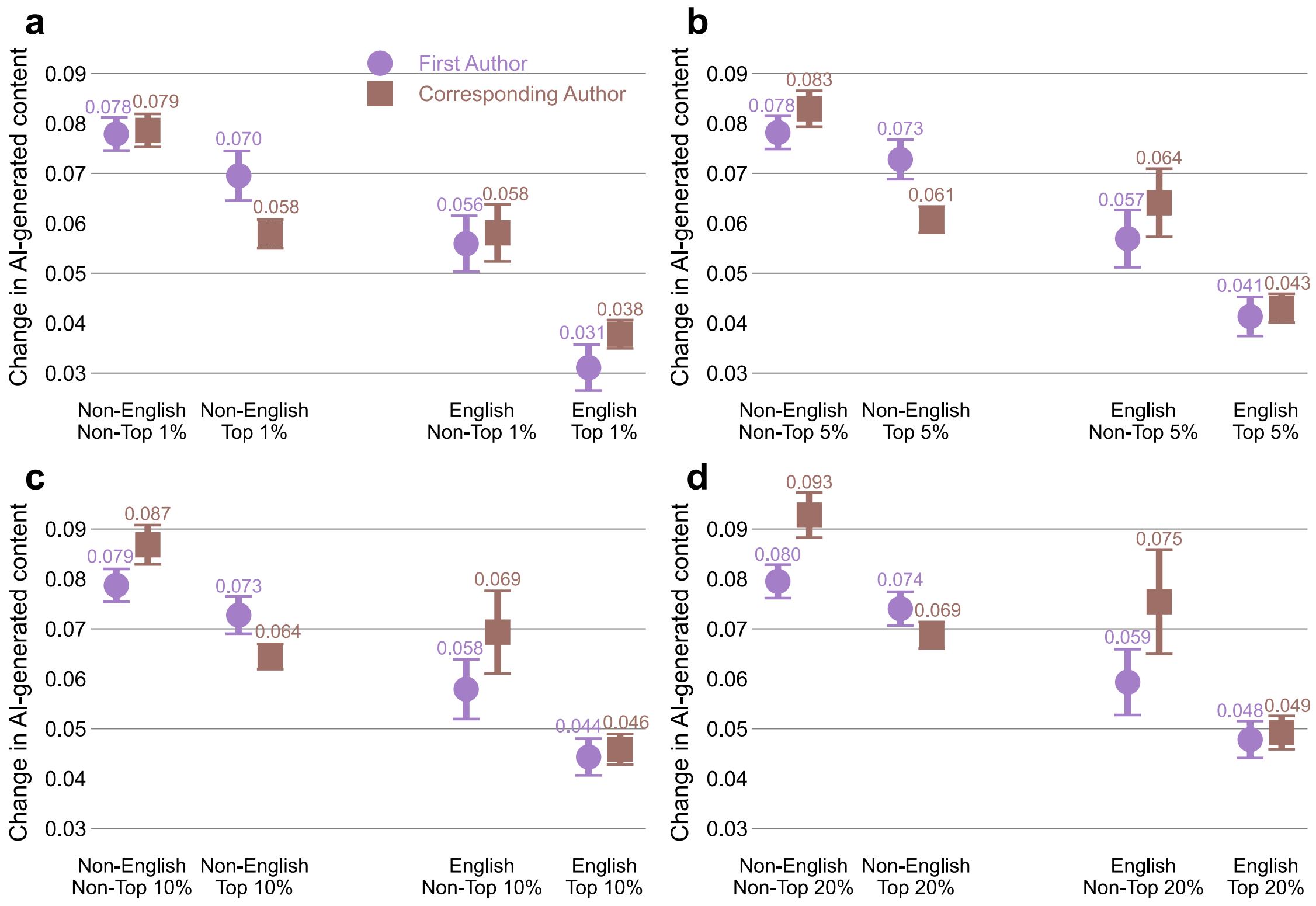

**Figure S4. Robustness check for the change in AI-generated content, stratifying authors by varying citation thresholds.** The figure shows the mean change in AI-generated content between pre- and post-ChatGPT periods, comparing authors from English- and non-English-speaking countries. Authors are designated as "top" or "non-top" based on their cumulative citation count. To test the robustness of this comparison, the analysis is repeated using four different thresholds to define the "top" group: **(a)** top 1%, **(b)** top 5%, **(c)** top 10%, and **(d)** top 20%. Results are shown separately for first authors (purple circles) and corresponding authors (brown squares). Values represent the change in the fraction of AI-generated content between pre- and post-ChatGPT periods, with 95% confidence intervals.

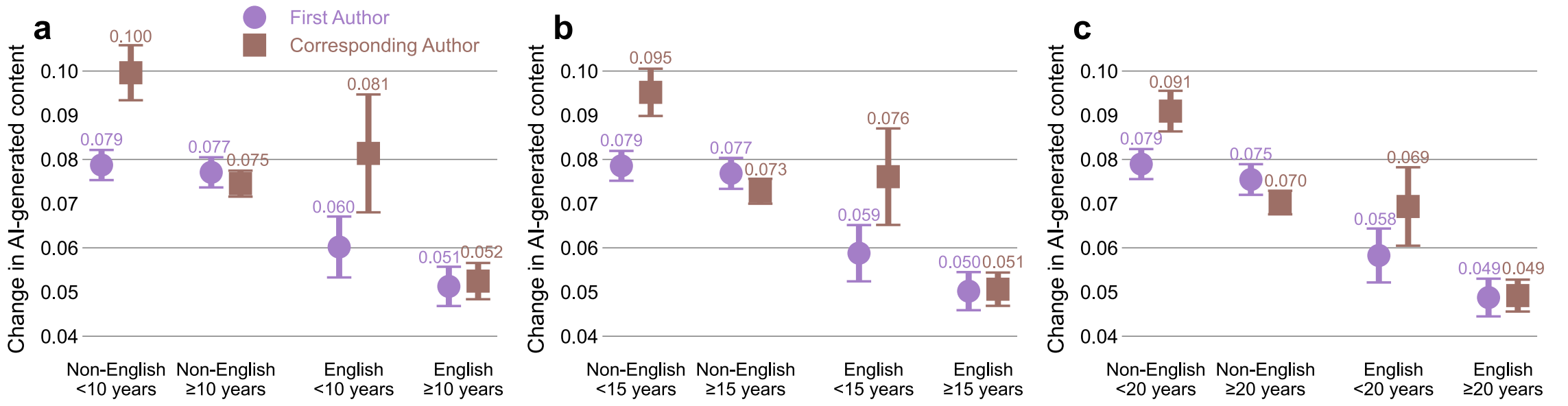

**Figure S5. Robustness check for the change in AI-generated content, stratifying authors by varying seniority thresholds.** The figure shows the mean change in AI-generated content between pre- and post-ChatGPT periods, comparing authors from English- and non-English-speaking countries. Authors are grouped by seniority based on their publication history (number of years). To test the robustness of this comparison, the analysis is repeated using three different seniority thresholds: **(a)** less than 10 years vs. 10 years or more; **(b)** less than 15 years vs. 15 years or more; **(c)** less than 20 years vs. 20 years or more. In each subplot, authors are grouped as "junior" (less than the threshold) or "senior" (greater than or equal to the threshold). Results are shown for first authors (purple circles) and corresponding authors (brown squares). Values represent the change in the fraction of AI-generated content between pre- and post-ChatGPT periods, with 95% confidence intervals.

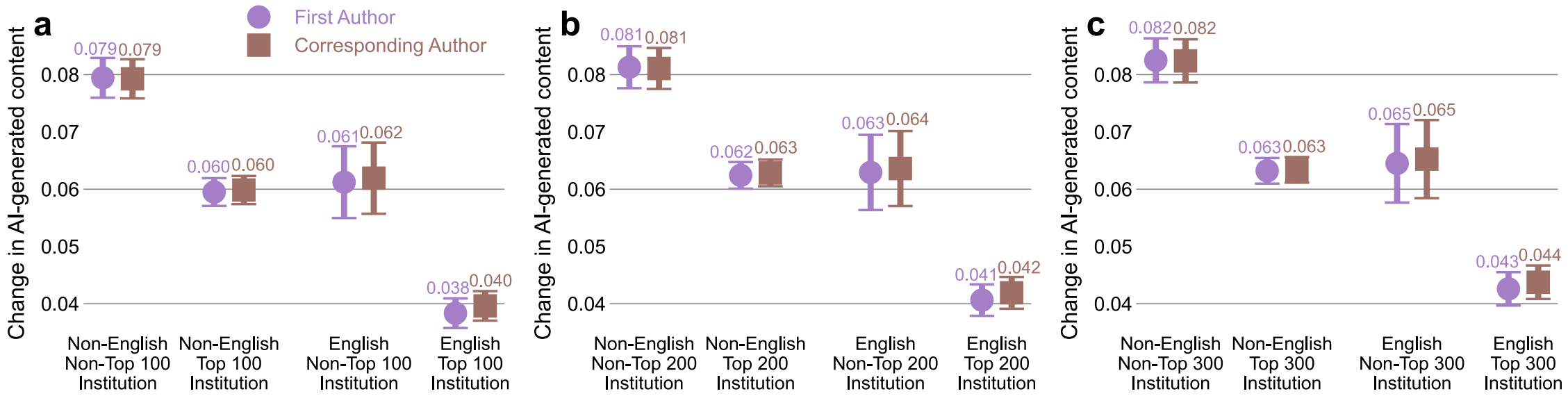

**Figure S6. Robustness check for the change in AI-generated content, stratifying authors by varying institution prestige thresholds.** The figure shows the mean change in AI-generated content between pre- and post-ChatGPT periods, comparing authors from English- and non-English-speaking countries. Authors are designated as "top" or "non-top" based on their institution's ranking. To test the robustness of this comparison, the analysis is repeated using three different ranking thresholds to define the "top" group: **(a)** top 100 vs. non-top 100 institutions, **(b)** top 200 vs. non-top 200 institutions, and **(c)** top 300 vs. non-top 300 institutions. Results are shown separately for first authors (purple circles) and corresponding authors (brown squares). Values represent the change in the fraction of AI-generated content between pre- and post-ChatGPT periods, with 95% confidence intervals.

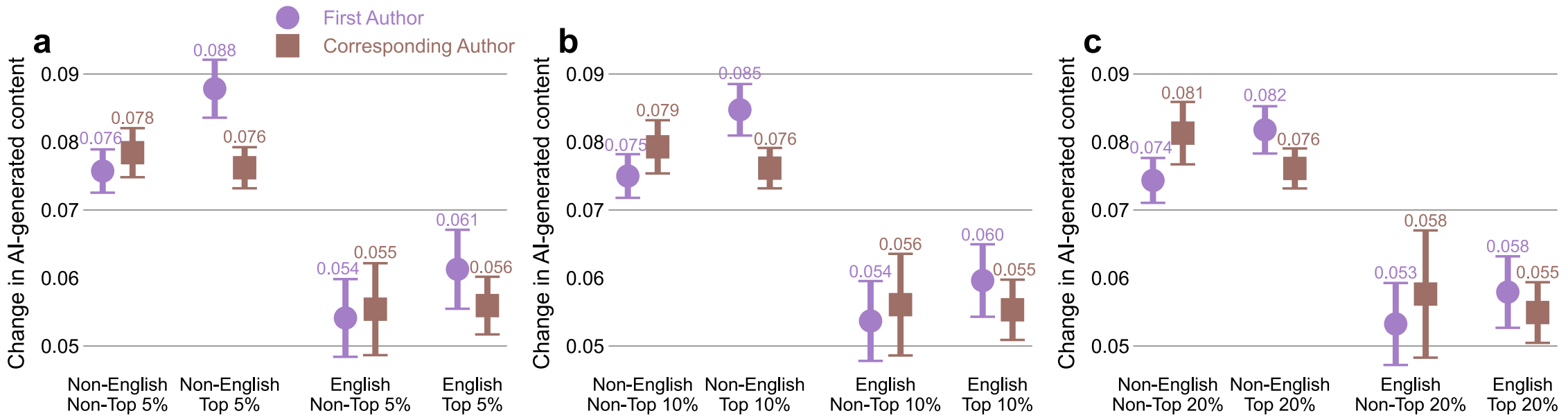

**Figure S7. Change in AI-generated content between pre- and post-ChatGPT periods, stratified by recent publication velocity and affiliation country group.** A publication was classified by whether the focal author was affiliated with an English- or non-English-speaking country, and by recent publication velocity (top vs. non-top). We measured recent publication velocity of a specific author as the number of an author's publications during the focal article's publication year and the immediately preceding year. Publication velocity was further normalized by year and subfield, and authors were classified into top and non-top groups using different thresholds (5%, 10%, and 20%). Values represent the change in the fraction of AI-generated content between pre- and post-ChatGPT periods, with 95% confidence intervals. For corresponding authors, the differences between the top and non-top groups were minimal. However, for first authors, papers by scientists with higher publication velocity in the recent two-year window showed a greater increase in AI-generated content than papers by their non-top counterparts, especially in non-English-speaking countries. This additional analysis suggests that first authors with higher recent publication velocity exhibited larger increases in estimated AI-assisted writing, particularly in non-English-speaking countries.

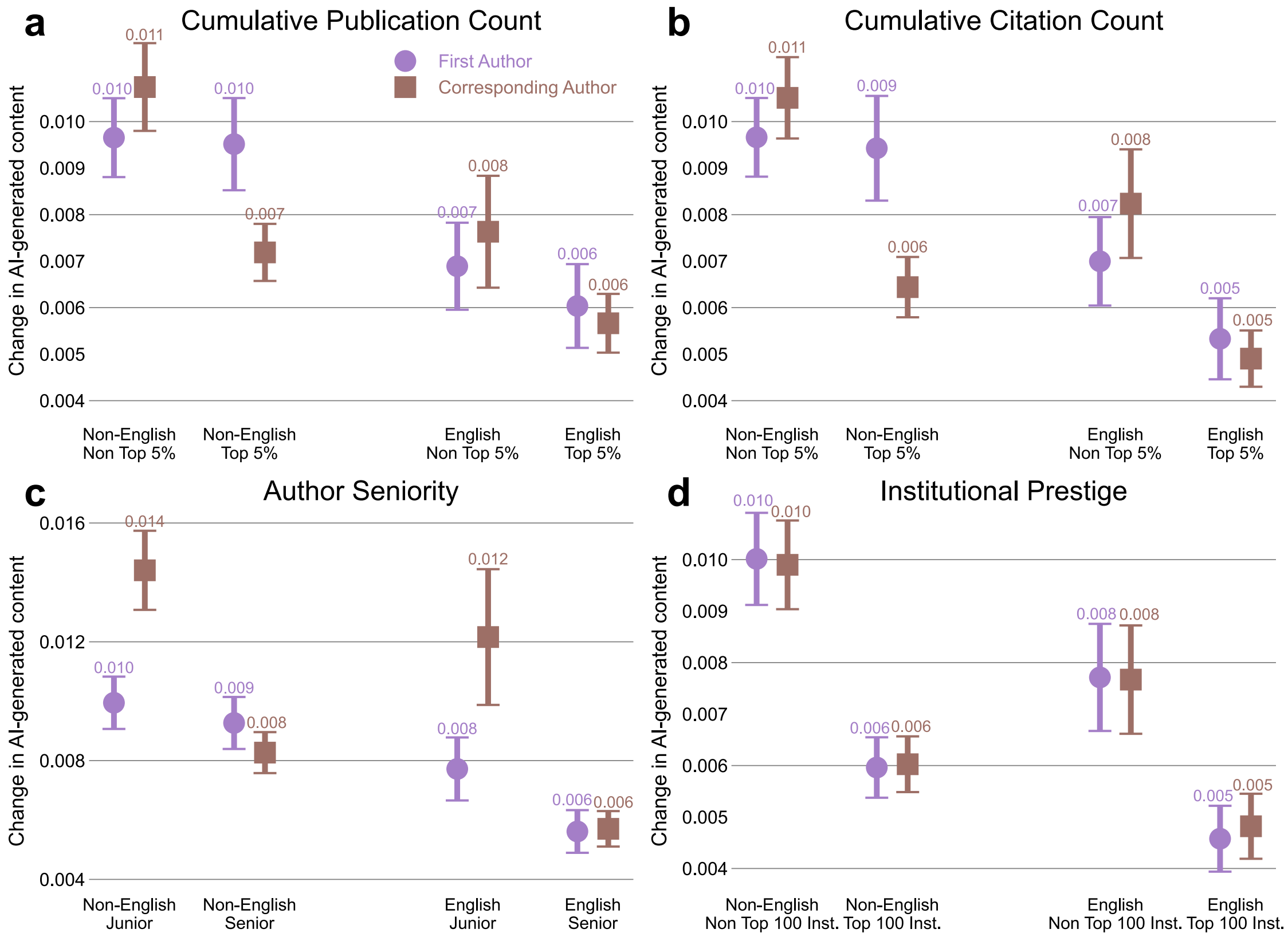

**Figure S8. Change in AI-generated content between pre- and post-ChatGPT periods, stratified by author profile and affiliation country group, based on estimations from the *Methods* and *Results* sections.** A publication was classified by whether the focal author was affiliated with an English- or non-English-speaking country, and by four author characteristics: **(a)** cumulative number of publications (top 5% vs. non-top 5%), **(b)** cumulative number of citations (top 5% vs. non-top 5%), **(c)** career seniority (senior vs. junior), and **(d)** institutional prestige (top 100 institutions vs. non-top 100). Values represent the change in the fraction of AI-generated content between pre- and post-ChatGPT periods, with 95% confidence intervals. Although the magnitude of the estimated increases is smaller than that observed in the *Introduction* and *Discussion* sections, the relative differences across linguistic contexts and author profiles are broadly consistent with the primary analyses.

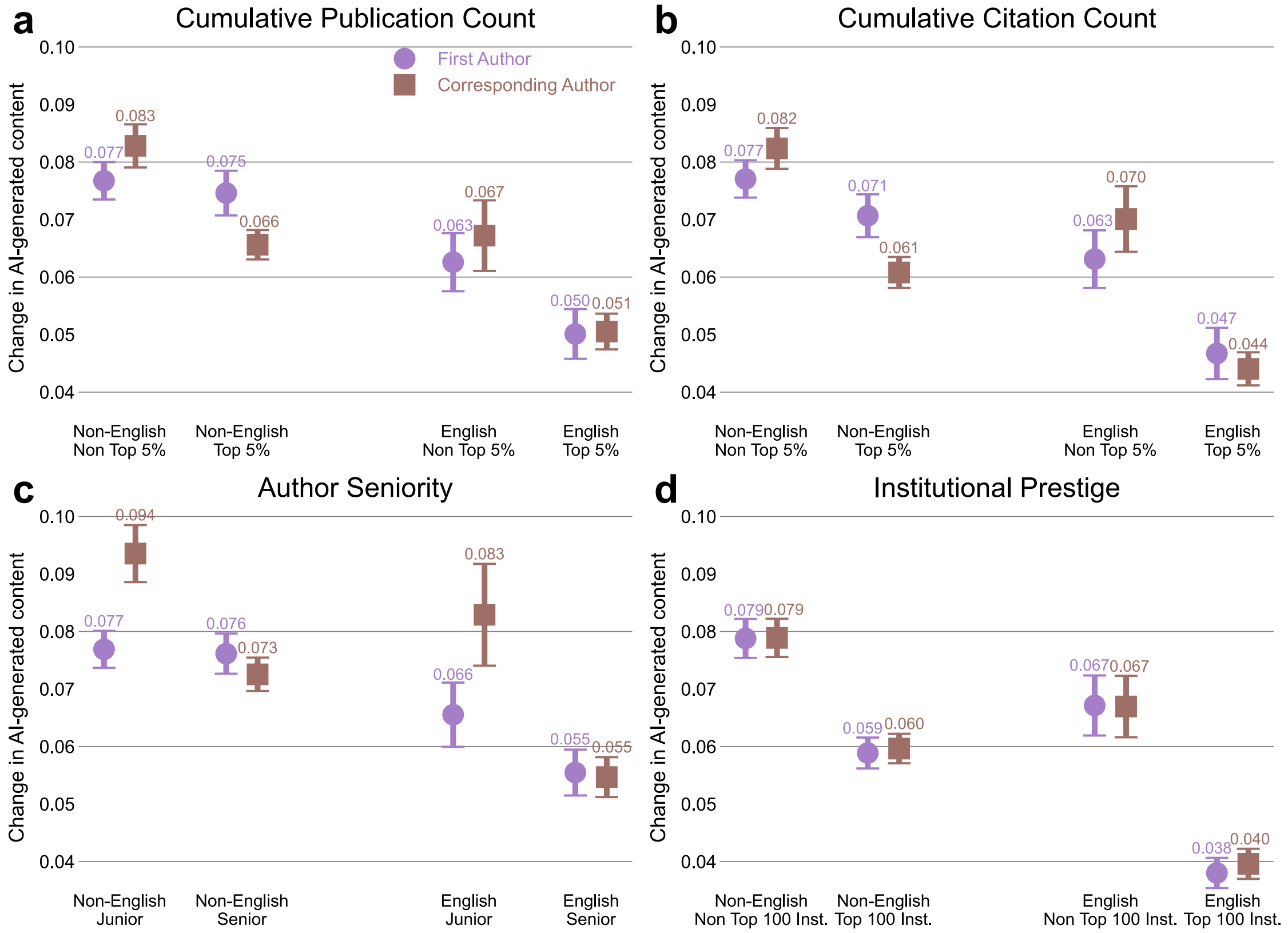

**Figure S9. Change in AI-generated content between pre- and post-ChatGPT periods, stratified by author profile and country group based on name-inferred nationality.** As a robustness check to the affiliation-based country classification, authors' likely nationalities were inferred from their names using name2nat, and the main analyses were repeated using this alternative classification. Results are shown separately according to the inferred nationality of first and corresponding authors and by four author characteristics: **(a)** cumulative number of publications (top 5% vs. non-top 5%), **(b)** cumulative number of citations (top 5% vs. non-top 5%), **(c)** career seniority (senior vs. junior), and **(d)** institutional prestige (top 100 institutions vs. non-top 100). Values represent the change in the fraction of AI-generated content between pre- and post-ChatGPT periods, with 95% confidence intervals.

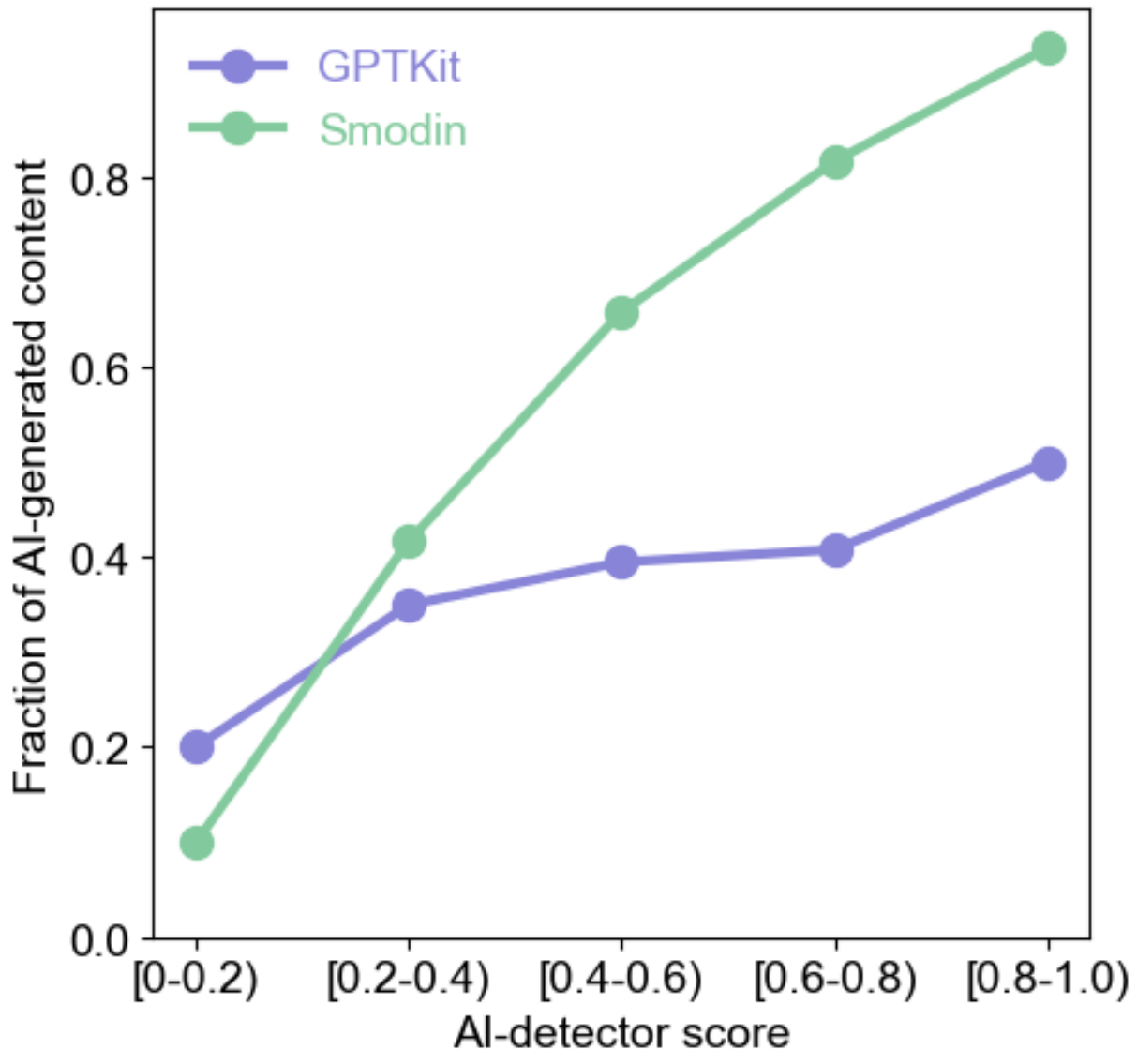

**Figure S10. Validation of the distribution-based strategy for estimating the fraction of AI-generated content**. To validate our estimation strategy, we compared its results with two AI detectors, GPTKit and Smodin. We utilized over 5,000 abstracts from our prior study [22] and grouped them by the AI-generated probability score provided by each detector (x-axis). We then applied our distribution-based strategy to estimate the fraction of AI-generated content for each group (y-axis). The positive associations for both detectors show correspondence between corpus-level estimates and the detector scores.

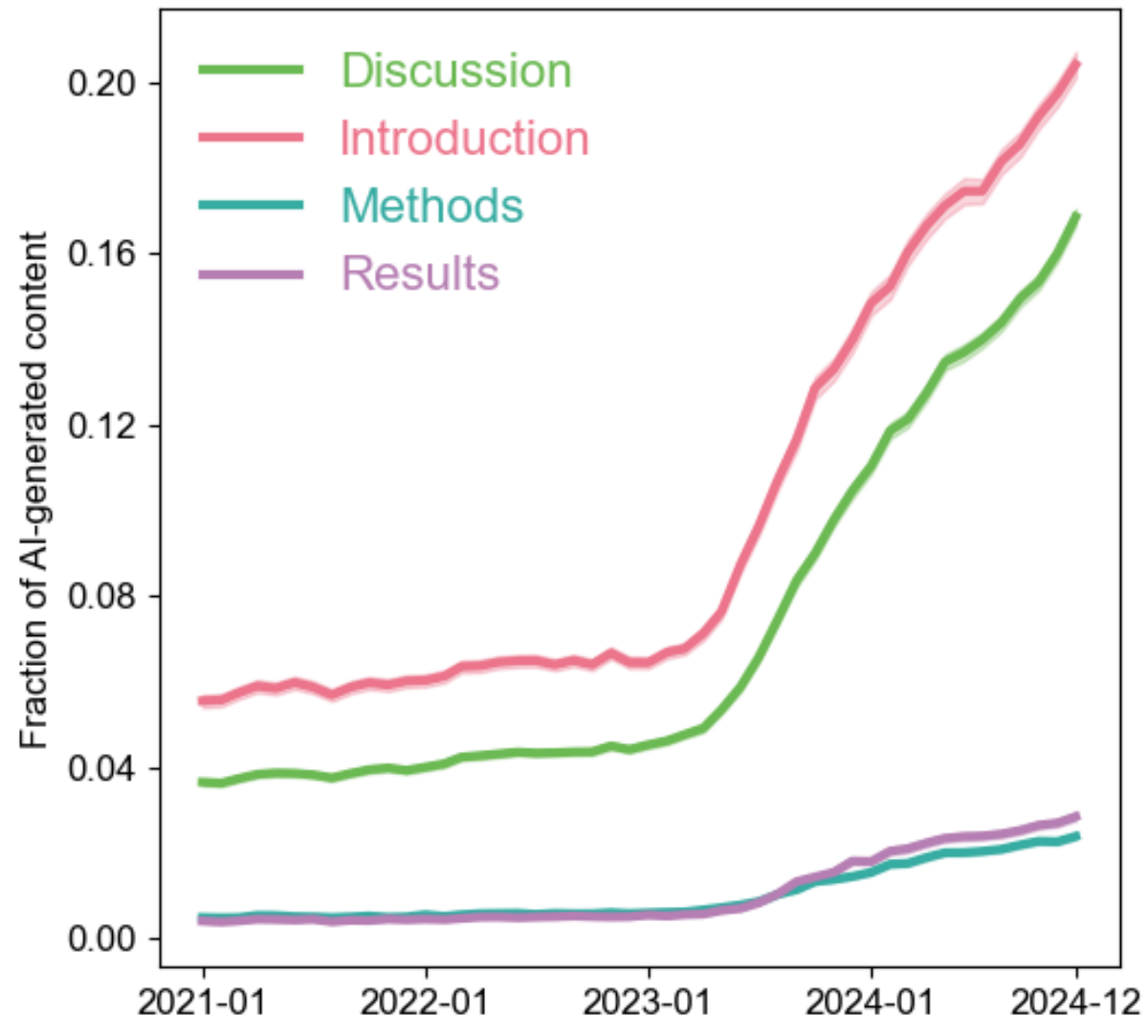

**Figure S11. Trends in the fraction of AI-generated content over time, by section of scientific publications (2021-2024).** Shaded regions denote the 95% confidence intervals. The post-ChatGPT increase is substantially larger in the *Introduction* and *Discussion* sections than in the *Methods* and *Results* sections, consistent with greater use of AI-assisted writing in more rhetorical and interpretive parts of scientific articles.

## Supplementary Tables

**Table S1. List of English-speaking countries/regions.**

| **English-speaking countries/regions where English is at least one of the languages spoken by at least 20% of the population** |
|---|
| Anguilla, Antigua and Barbuda, Aruba, Australia, Bahamas, Barbados, Belize, Bermuda, Botswana, Cameroon, Canada, Cayman Islands, Dominica, Egypt, Eritrea, Fiji, Gambia, Gibraltar, Great Britain, Grenada, Guyana, Hong Kong, India, Ireland, Israel, Jamaica, Jordan, Kenya, Kiribati, Kuwait, Lebanon, Lesotho, Liberia, Malta, Micronesia, New Zealand, Nigeria, Pakistan, Philippines, Puerto Rico, Rwanda, Saint Kitts and Nevis, Saint Lucia, Saint Vincent and the Grenadines, Samoa, Seychelles, Sierra Leone, Singapore, South Africa, South Korea, Trinidad and Tobago, United States, Vanuatu, Zambia, Zimbabwe. |

**Table S2. Difference-in-differences results.** The dependent variable is a paper's estimated fraction of AI-generated sentences. Standard errors clustered at the journal level are reported in parentheses. *** $p < .001$, ** $p < .01$, * $p < .05$.

| | (1)<br>First | (2)<br>First | (3)<br>Corresponding | (4)<br>Corresponding |
|---|---|---|---|---|
| EnglishSpeakingCountry | 0.019*** | 0.017*** | 0.019*** | 0.016*** |
| | (0.002) | (0.001) | (0.002) | (0.001) |
| PostGPT | 0.078*** | 0.078*** | 0.078*** | 0.078*** |
| | (0.003) | (0.002) | (0.003) | (0.002) |
| EnglishSpeakingCountry*PostGPT | -0.023*** | -0.023*** | -0.022*** | -0.022*** |
| | (0.003) | (0.003) | (0.003) | (0.003) |
| Log(AuthorCounts) | -0.020*** | -0.006*** | -0.019*** | -0.005*** |
| | (0.002) | (0.001) | (0.002) | (0.001) |
| Log(RefCounts) | 0.027*** | 0.034*** | 0.027*** | 0.034*** |
| | (0.002) | (0.001) | (0.002) | (0.001) |
| Log(NumSentences) | -0.011*** | -0.032*** | -0.011*** | -0.032*** |
| | (0.003) | (0.002) | (0.003) | (0.002) |
| Journal Fixed Effect | No | Yes | No | Yes |
| Subfield Fixed Effect | No | Yes | No | Yes |
| Observations | 1,979,747 | 1,978,736 | 1,951,427 | 1,950,463 |
| R-squared | 0.084 | 0.193 | 0.084 | 0.192 |

**Table S3. Difference-in-difference-in-differences results.** The dependent variable is a paper's estimated fraction of AI-generated sentences. In models (1) and (2), AuthorProfileGroup = 1 represents papers by first/corresponding authors ranked top 5% in their subfield in that year in terms of the number of cumulative publications; otherwise AuthorProfileGroup = 0. In models (3) and (4), AuthorProfileGroup = 1 represents papers by first/corresponding authors ranked top 5% in their subfield in that year in terms of the number of cumulative citations; otherwise AuthorProfileGroup = 0. In models (5) and (6), AuthorProfileGroup = 1 represents papers authored by senior scientists; otherwise AuthorProfileGroup = 0. In models (7) and (8), AuthorProfileGroup = 1 represents papers authored by scientists from top 100 institutions based on the 2025 QS World University Rankings in Life Sciences and Medicine; otherwise, AuthorProfileGroup = 0. Standard errors clustered at the journal level are reported in parentheses. *** $p < .001$, ** $p < .01$, * $p < .05$.

| | (1) Publication First | (2) Publication Corresponding | (3) Citation First | (4) Citation Corresponding | (5) Seniority First | (6) Seniority Corresponding | (7) Institution First | (8) Institution Corresponding |
|---|---|---|---|---|---|---|---|---|
| EnglishSpeakingCountry | 0.017*** | 0.016*** | 0.017*** | 0.016*** | 0.017*** | 0.018*** | 0.015*** | 0.014*** |
| | (0.001) | (0.002) | (0.001) | (0.001) | (0.001) | (0.002) | (0.001) | (0.001) |
| PostGPT | 0.078*** | 0.084*** | 0.078*** | 0.083*** | 0.079*** | 0.095*** | 0.079*** | 0.079*** |
| | (0.002) | (0.002) | (0.002) | (0.002) | (0.002) | (0.002) | (0.002) | (0.002) |
| AuthorProfileGroup | -0.0002 | -0.0002 | -0.001** | -0.0001 | -0.001*** | -0.002*** | 0.002*** | 0.002*** |
| | (0.0004) | (0.0005) | (0.0005) | (0.001) | (0.0003) | (0.001) | (0.001) | (0.001) |
| PostGPT*EnglishSpeakingCountry | -0.022*** | -0.023*** | -0.021*** | -0.019*** | -0.020*** | -0.019*** | -0.018*** | -0.017*** |
| | (0.003) | (0.003) | (0.003) | (0.003) | (0.003) | (0.005) | (0.003) | (0.003) |
| PostGPT*AuthorProfileGroup | -0.002 | -0.018*** | -0.005*** | -0.022*** | -0.002 | -0.022*** | -0.020*** | -0.019*** |
| | (0.001) | (0.001) | (0.001) | (0.001) | (0.001) | (0.002) | (0.002) | (0.002) |
| EnglishSpeakingCountry*AuthorProfileGroup | -0.002* | -0.001 | -0.001 | 0.001 | -0.002* | -0.002 | 0.004*** | 0.004*** |
| | (0.001) | (0.001) | (0.001) | (0.001) | (0.001) | (0.002) | (0.001) | (0.001) |
| PostGPT*EnglishSpeakingCountry*AuthorProfileGroup | -0.010*** | 0.007* | -0.010*** | 0.001 | -0.007*** | -0.003 | -0.003 | -0.003 |
| | (0.002) | (0.003) | (0.002) | (0.003) | (0.001) | (0.004) | (0.003) | (0.003) |
| Log(AuthorCounts) | -0.006*** | -0.005*** | -0.006*** | -0.005*** | -0.006*** | -0.005*** | -0.006*** | -0.005*** |
| | (0.001) | (0.001) | (0.001) | (0.001) | (0.001) | (0.001) | (0.001) | (0.001) |
| Log(RefCounts) | 0.034*** | 0.035*** | 0.034*** | 0.035*** | 0.034*** | 0.035*** | 0.034*** | 0.035*** |
| | (0.001) | (0.001) | (0.001) | (0.001) | (0.001) | (0.001) | (0.001) | (0.001) |
| Log(NumSentences) | -0.032*** | -0.032*** | -0.032*** | -0.032*** | -0.032*** | -0.032*** | -0.032*** | -0.032*** |
| | (0.002) | (0.002) | (0.002) | (0.002) | (0.002) | (0.002) | (0.002) | (0.002) |
| Journal Fixed Effect | Yes | Yes | Yes | Yes | Yes | Yes | Yes | Yes |
| Subfield Fixed Effect | Yes | Yes | Yes | Yes | Yes | Yes | Yes | Yes |
| Observations | 1,978,736 | 1,950,463 | 1,978,736 | 1,950,463 | 1,954,819 | 1,950,463 | 1,962,592 | 1,919,166 |
| R-squared | 0.193 | 0.194 | 0.193 | 0.195 | 0.195 | 0.195 | 0.194 | 0.193 |

**Table S4. DiD analysis with author fixed effect.** The dependent variable is a paper's estimated fraction of AI-generated sentences. Standard errors clustered at the journal level are reported in parentheses. *** $p < .001$, ** $p < .01$, * $p < .05$.

| | (1)<br>First | (2)<br>Corresponding |
|---|---|---|
| PostGPT | 0.066*** | 0.065*** |
| | (0.001) | (0.001) |
| EnglishSpeakingCountry*PostGPT | -0.020*** | -0.018*** |
| | (0.002) | (0.002) |
| Log(AuthorCounts) | -0.008*** | -0.007*** |
| | (0.001) | (0.001) |
| Log(RefCounts) | 0.029*** | 0.031*** |
| | (0.001) | (0.001) |
| Log(NumSentences) | -0.030*** | -0.033*** |
| | (0.002) | (0.001) |
| Journal Fixed Effect | Yes | Yes |
| Subfield Fixed Effect | Yes | Yes |
| Author Fixed Effect | Yes | Yes |
| Observations | 965,125 | 1,391,679 |
| R-squared | 0.615 | 0.499 |

**Table S5. Author-level OLS linear regression results for changes in AI-generated content.** The dependent variable is the change in the estimated fraction of AI-generated content for an author. Robust standard errors are used and reported in parentheses. *** $p < .001$, ** $p < .01$, * $p < .05$.

| | (1) First | (2) First | (3) First | (4) Corresponding | (5) Corresponding | (6) Corresponding |
|---|---|---|---|---|---|---|
| EnglishSpeakingCountry | -0.021*** | -0.021*** | -0.019*** | -0.022*** | -0.022*** | -0.020*** |
| | (0.001) | (0.001) | (0.001) | (0.001) | (0.001) | (0.001) |
| AI_author | | 0.013*** | 0.016*** | | 0.011*** | 0.012*** |
| | | (0.001) | (0.001) | | (0.001) | (0.001) |
| EnglishSpeakingCountry*AI_author | | | -0.013*** | | | -0.005* |
| | | | (0.002) | | | (0.002) |
| Log(CumulativePubs) | 0.011*** | 0.009*** | 0.009*** | 0.012*** | 0.010*** | 0.010*** |
| | (0.001) | (0.001) | (0.001) | (0.001) | (0.001) | (0.001) |
| Log(CumulativeCits) | -0.004*** | -0.004*** | -0.003*** | -0.008*** | -0.008*** | -0.008*** |
| | (0.0004) | (0.0004) | (0.0004) | (0.0003) | (0.0003) | (0.0003) |
| CareerAge | -0.0004*** | -0.0004*** | -0.0004*** | -0.001*** | -0.001*** | -0.001*** |
| | (0.000) | (0.000) | (0.000) | (0.000) | (0.000) | (0.0000) |
| TopInstitution | -0.023*** | -0.023*** | -0.023*** | -0.022*** | -0.022*** | -0.022*** |
| | (0.001) | (0.001) | (0.001) | (0.001) | (0.001) | (0.001) |
| Author Subfield Fixed Effect | Yes | Yes | Yes | Yes | Yes | Yes |
| Observations | 163,045 | 163,045 | 163,045 | 200,846 | 200,846 | 200,846 |
| R-squared | 0.018 | 0.018 | 0.019 | 0.023 | 0.024 | 0.024 |

**Table S6. Author-level OLS linear regression results for changes in productivity.** The dependent variable is the change in the number of publications for an author between the pre- and post-ChatGPT periods. Robust standard errors are used and reported in parentheses. *** $p < .001$, ** $p < .01$, * $p < .05$.

| | (1)<br>First | (2)<br>First | (3)<br>Corresponding | (4)<br>Corresponding |
|---|---|---|---|---|
| LLM_use_change | 2.456*** | 2.558*** | 3.334*** | 3.458*** |
| | (0.052) | (0.060) | (0.050) | (0.060) |
| EnglishSpeakingCountry | 0.224*** | 0.246*** | 0.468*** | 0.496*** |
| | (0.020) | (0.021) | (0.021) | (0.022) |
| EnglishSpeakingCountry*LLM_use_change | | -0.382*** | | -0.400*** |
| | | (0.116) | | (0.108) |
| Log(CumulativePubs) | -0.428*** | -0.429*** | -0.828*** | -0.828*** |
| | (0.013) | (0.013) | (0.014) | (0.014) |
| Log(CumulativeCits) | -0.013 | -0.013 | 0.121*** | 0.121*** |
| | (0.008) | (0.008) | (0.007) | (0.007) |
| CareerAge | 0.009*** | 0.009*** | 0.007*** | 0.007*** |
| | (0.001) | (0.001) | (0.001) | (0.001) |
| TopInstitution | 0.008 | 0.005 | 0.071** | 0.069** |
| | (0.026) | (0.026) | (0.026) | (0.026) |
| Author Subfield Fixed Effect | Yes | Yes | Yes | Yes |
| Observations | 163,045 | 163,045 | 200,846 | 200,846 |
| R-squared | 0.035 | 0.035 | 0.051 | 0.051 |

## Supplementary References